\documentclass[preprint,tighten,floats,twoside,eqsecnum,aps,epsf,epsfig]{revtex4}
\usepackage{graphicx}

\usepackage{color}
\usepackage{bm}% bold math
\newcommand{\GeV}{\makebox{ GeV}}
\newcommand{\beq}{\begin{equation}}
\newcommand{\enq}{\end{equation}}
\newcommand{\beqa}{\begin{eqnarray}}
\newcommand{\beqast}{\begin{eqnarray*}}
\newcommand{\enqa}{\end{eqnarray}}
\newcommand{\enqast}{\end{eqnarray*}}

\def\GeV{\nobreak\,\mbox{GeV}}

\begin{document}
     %  \input{part1}
     %  \input{conclusions}
     %  \input{appendix}
     %  \input{liter}
% ===========================
\title{Amplitudes in the Coulomb \\
interference region of pp and p$\bar {\rm p} $  scattering  }
 
  \author{Anderson Kendi, Erasmo Ferreira and Takeshi Kodama } 
 \affiliation{Instituto de F\'{\i}sica, Universidade Federal do Rio de
 Janeiro \\
 C.P. 68528, Rio de Janeiro 21945-970, RJ, Brazil   }

\begin{abstract}
 
   We discuss the determination of the parameters of the pp and 
 p$\bar{\rm p} $ amplitudes for the description of scattering 
in the Coulomb interference region. We put enphasis on the possibility
that the effective slope observed in the differential cross section 
is formed by different exponential slopes in the real and 
imaginary amplitudes (called $B_R$ and $B_I$).  For this purpose we 
develop  a more general treatment of the Coulomb phase.
We analyse the differential cross section data in the range from 19 
to 1800 GeV with four parameters ($\sigma ~ , ~ \rho ~ , ~ B_I ~, ~B_R $ ),
and observe that we cannot obtain from the data a unique determination 
of the parameters. 
We investigate correlations in pairs of the four quantities, showing 
ranges leading to the smaller $\chi^2$ values. 
      
In the specific case of  p$\bar {\rm p} $  scattering at 541 GeV, 
 we investigate the measurements of event rate dN/dt
at low $|t|$ \cite{augier} in terms of the  Coulomb interference with 
exponentially decreasing nuclear amplitudes. The  
analysis allows  a  determination of the normalization factor
connecting the event rate with the absolute cross section.
\end{abstract}

\bigskip
  \maketitle

\section{Introduction}\label{intro}

The experiments in the high energy accelerators of Cern and Fermilab 
in the period from 1960 to 1990 collected data on the differential 
cross sections  $d\sigma/dt$ for the systems pp and p$\bar {\rm p} $
at the center of mass energies $\sqrt{s}=20-1800$ GeV . 
Along almost 50 years since the beginning of these studies of 
high energy hadronic scattering, many theoretical models were 
developed, but the experimental data stopped increasing in quantity or 
quality, as the accelerators were discontinued. The phenomenology 
and theoretical treatments of these systems are  thus restricted in
several aspects. The theoretical literature is enormous, now 
with increased interest due to the higher energy data that will 
come from LHC operation  \cite{fiore}, 
and in the present work we do not analyse or compare the many 
dynamical and phenomenological efforts.

To study of the dynamics that governs the processes, it is necessary to 
disentangle the squared moduli  of complex quantities that represent  
the measured quantities  in terms of the imaginary and real parts, 
 which are intrinsically combined with the Coulomb contribution. 

In the region of small momentum transfers $|t|$, the intense Coulomb 
amplitude  added to the nuclear 
interaction  creates an interference that is observable in the 
$|t|$ distribution in $d\sigma/dt$. This Coulomb interference region 
of low $|t|$ values  goes typically  up to $|t|= 0.01 ~ {\rm GeV}^2$, 
but we show that the form 
of $d\sigma/dt$ in general can actually be described in terms of  simple  
exponential real and imaginary nuclear amplitudes well beyond this range. 

 In previous analysis of the pp and p$\bar {\rm p} $ data, the real and 
imaginary nuclear amplitudes were considered as having the same exponential 
dependence  $\exp {(Bt/2)}$ , where $B$ is the slope of the log plot of
$d\sigma/dt$ . This simplifying assumption  is not adequate, according 
to dispersion relations \cite{EF2007} and according to the theorem of 
A. Martin \cite{Martin} that says that the position of the zero of the real 
amplitude is close and approaches $t=0$  as the energy increases.
Both results indicate that the slope of the real amplitude should be 
larger than that of the imaginary one, and in the present work 
we investigate the description of the Coulomb interference region allowing 
 for different real and imaginary  slopes. We review the scattering data 
in cases where this kind of information can be looked for.

In Sec. II we review the expressions for the observable  quantities in the 
forward region, and obtain the  expression of the intervening 
relative phase for the more general case of different  slopes for the real 
and imaginary  amplitudes. In Sec. III we analyse the differential cross 
sections for the energies in pp and   p$\bar {\rm p} $  scattering  
where data are more favorable. In Sec. IV we present some remarks and 
conclusions. 

\section {Low $|t|$ region and Coulomb phase}

\subsection{Description of scattering for small $|t|$ }
In elastic pp and p$\bar {\rm p} $  collisions, the combined nuclear 
and coulomb amplitudes   is written 
\begin{equation}
\label{complete_amplitude}
F^{C+N}(s,t)= F^C(s,t) e^{i \alpha \Phi (s,t)} + F^N (s,t) ~ , 
\end{equation} 
where   $F^C$  is the Coulomb part
\begin{equation}
\label{coulomb}
F^C= (-/+)~ \frac{2 \alpha}{|t|} ~  F^2_{\rm proton}
\end{equation}
with the proton electromagnetic form factor
\begin{equation}
\label{ff_proton}
F_{\rm proton}=(0.71/(0.71+|t|))^2 ~ , 
\end{equation}
associated to a relative phase 
$\Phi$, and $F^N$ is the strong interaction complexe amplitude 
\begin{equation}
\label{nuclear}
F^N (s,t)=F^N_R(s,t) + i ~ F^N_I (s,t) ~ . 
\end{equation}

The phase $\Phi$ was initially studied by West  and Yennie \cite{WY}, 
and different evaluations have been worked out by several authors 
\cite{selyugin,petrov,KL}. In the present work  we extend these 
investigations considering the possibility of different slopes for the 
real and imaginary amplitudes. 

In the normalization that we use \cite{nicolescu} the differential 
cross section is written
\begin{equation}
\label{dsigdt1}
\frac{d\sigma}{dt}= \pi |F^{C+N}(s,t)|^2=
  \pi |F^C(s,t)e^{i \alpha \Phi (s,t)}+ F^N_R(s,t) + i F^N_I(s,t)|^2 ~ .
\end{equation}
For small angles we can approximate
\begin{equation}
\label{small_t}
F^N(s,t)\approx F^N_R(s,0) e^{B_R t/2}+ i F^N_I(s,0) e^{B_I t/2} ~ .
\end{equation}

  The slopes $B_R$ and $B_I$ are usually treated as having equal 
values. In the present work we allow $B_R \neq B_I$.

The parameter
\begin{equation}
\label{rho}
\rho=\frac{F^N_R(s,0)}{F^N_I(s,0)}  ~ , 
\end{equation}
the optical theorem
\begin{equation}
\label{optical}
\sigma= 4 \pi ~ (0.389) ~ {\rm Im}~ {F^N_I(s,0)} ~ , 
\end{equation}
and the slopes $B_R$, $B_I$ are used to parametrize the 
differential cross section for small $|t|$ .  In these 
expressions, $\sigma$ is in milibarns   and the amplitudes 
$F_R$, $F_I$ are in $\GeV ^{-2}$ .  

For low $|t|$ , eq. (\ref{small_t}) leads to the approximate form 
\begin{equation}
\label{sig_forward}
\frac{d\sigma}{dt}=\bigg| \frac{d\sigma}{dt}\bigg|_{t=0} ~  e^{Bt}  ~ , 
\end{equation}
with
\begin{equation}
\label{global_slope}
B=\frac{\rho^2 B_R +B_I}{1+\rho^2}  ~  
\end{equation}
as the usual slope observed in the data of $d{\sigma}/dt$. 

\subsection {The Coulomb phase} 

Here we derive an expression for the phase appropriate for cases 
with $B_R \neq B_I$. 

The starting point is the expression for the phase obtained by West and Yennie
\cite{WY}
\begin{equation} 
\label{WYphase}
\Phi(s,t)=(-/+)\Bigg[\ln\bigg(-\frac{t}{s}\bigg)+\int_{-4p^2}^{0}
\frac{dt^\prime}{|t^\prime-t|}\bigg[1-\frac{F^N(s,t^\prime)}{F^N(s,t)}\bigg] ~ , 
\Bigg]
\end{equation}
where the signs $(-/+)$ are applied to the choices pp/p$\bar {\rm p} $ 
respectively.
The quantity $p$ is the proton momentum im center of mass system,
and at high energies $ 4p^2 \approx s $ . 

  For small $|t|$, assuming that $F^N(s,t^\prime)$ keeps the 
same form for large $|t^\prime|$ (this approximation should 
not have practical importance for the results), we have 
\begin{eqnarray}
\label{relation1}
\frac{F^N(s,t^\prime)}{F^N(s,t)}&=& \frac{F_R^N(s,0) e^{B_Rt^\prime/2}
+i ~ F_I^N(s,0) e^{B_It^\prime/2}}
{F_R^N(s,0) e^{B_Rt/2}
+i ~ F_I^N(s,0) e^{B_It/2}} \nonumber \\ 
&=&\frac{c}{c+i} ~ e^{B_R(t^\prime-t)/2}
+\frac{i } {c+i} ~  e^{B_I(t^\prime-t)/2} ~, 
\end{eqnarray}
where  
\begin{equation}
\label{def_c}
c \equiv  \rho e^{(B_R-B_I)t/2}   ~ .  
\end{equation}
The calculation is explained in detail in the appendix. 
The integrals  that appear in the evaluation of  eq. (\ref{WYphase}) are reduced to the  form \cite{KL} 
 \begin{equation}
\label{int_form}
I(B)=\int_{-4p^2}^{0}
\frac{dt^\prime}{|t^\prime-t|}\bigg[1-e^{B(t^\prime-t)/2}\bigg] ~  
\end{equation}
that is solved in terms of exponential integrals \cite{abramowitz} as 
\begin{equation}
\label{functional_form}
I(B)=   E_1\big[\frac{B}{2}\bigg(4p^2+t\bigg)\big] 
-E_i \big[-\frac{Bt}{2} \big] +\ln  \big[\frac{B}{2}\bigg(4p^2+t\bigg)\big]
-\ln  \big[-\frac{Bt}{2} \big] + 2 \gamma  ~ . 
\end{equation}  
 
The real part of the phase is then written 
\begin{equation}
\label{realphase}
  \Phi(s,t)  =(-/+)\Bigg[\ln \bigg(-\frac{t}{s} \bigg) 
+ \frac{1}{c^2+1}\bigg[ c^2 I(B_R)+I(B_I) \bigg]  
\Bigg]  ~ ,     
\end{equation}
and this expression in introduced into eq. ( \ref{dsigdt1} ) . 
 
%  \clearpage 

\section{Analysis of experimental data}

With $\sigma$ in mb  and  $t$ in GeV$^2$ the practical 
expession for $d\sigma / dt$ in terms of the parameters
$\sigma$, $\rho$ , $B_I$ and $B_R$ is 
\begin{equation}
\label{dsigdt}
\frac{d\sigma}{dt}=0.389 ~ \pi \Bigg[
\bigg[ \frac{\rho ~ \sigma ~ e^{B_R t /2}  } {0.389 \times 4 \pi}
   + F^C \cos(\Phi) \bigg]^2
+\bigg[  \frac{\sigma ~ e^{B_I t /2} } {0.389 \times 4 \pi}
+ F^C \sin(\Phi) \bigg]^2
\Bigg]   ~ , 
\end{equation}  
where by $\Phi$ we mean the real part given in eq. (\ref{realphase}),
written 
\begin{equation} 
\label{sphase}
\Phi=(-/+) ~ \alpha ~ \Bigg[ \ln \bigg(-\frac{t}{s}\bigg)+ Z_R \Bigg] ~ , 
\end{equation}
where 
\begin{equation}
Z_R= \frac{1}{1+c^2} \bigg[c^2 I(B_R) + I(B_I) \bigg] ~ ,  
\end{equation} 
with $c$ given in Eq. (\ref{def_c}).

At high energies and small $|t|$ we simplify
$$ 4 p^2 + t \rightarrow s $$ 
and then the functional form of $I(B)$ is written  
\begin{equation}
\label{functional_form}
I(B)=   E_1\bigg(\frac{B s }{2} \bigg) 
-E_i \bigg(-\frac{Bt}{2} \bigg) +\ln  \bigg(\frac{Bs}{2} \bigg)
-\ln  \bigg(-\frac{Bt}{2} \bigg) + 2 \gamma   ~ .  
\end{equation}
 
We have used  fitting programs (Cern Minuit-PAW and Numerical Recipes)
 to obtain correlations for the four parameters 
( $  \sigma ,  \rho , B_R ,  B_I $ ), for some values of energy where 
the data from CERN and Fermilab \cite{databasis} have more quality 
and quantity. Below we present these cases. 

It is important to remark that the results obtained in the fittings 
in general depend strongly on the set of data of low $|t|$ selected for the
analysis of the Coulomb interference region.  This shows that 
the  data   accumulated  in these experiments are not detailed 
and regular enough to allow precise determination of the amplitudes
in the forward direction. 

We stress that in this paper  we do not intend to give new better values 
for parameters. Instead, we show that the analysis of the data leads to 
rather ample possibilities.  

The values of forward scattering parameters for pp scattering at Fermilab and 
Cern ISR energies given in the standard literature  are given in Table \ref{basic}.
The data at $\sqrt{s}=19.4$ GeV come from Fermilab , and 
that at 23.5 - 62.5 GeV are from the  Cern ISR, with a review by 
Amaldi and Schubert \cite{amaldi}. 

\begin{center}
   \begin{table}
   \caption{ \label{basic} Forward scattering parameters found in the literature.
    At 1800 GeV the data values  $1800^{(a)}$ and $1800^{(b)}$ correspond to 
   the experiments E710 and E741 in Fermilab}
   \vspace{.5 cm}
   \begin{tabular}{ |c|c|c|c| }    
   \hline 
$\sqrt{s}$ (GeV) & $\sigma $ (mb) & $ \rho $ & $B ({\rm GeV}^{-2})$  \\
 \hline
19.4 &  $38.98 \pm 0.04 $ & $0.019  \pm 0.016$ & $11.74 \pm 0.04$  \\ 
23.5 &  $38.94 \pm 0.17 $ & $0.02   \pm 0.05 $ & $11.80 \pm 0.30$  \\
30.7 &  $40.14 \pm 0.17 $ & $0.042  \pm 0.011$ & $12.20 \pm 0.30$  \\
44.7 &  $41.79 \pm 0.16 $ & $0.0620 \pm 0.011$ & $12.80 \pm 0.20$  \\ 
52.8 &  $42.67 \pm 0.19 $ & $0.078  \pm 0.010$ & $12.87 \pm 0.14$  \\ 
62.5 &  $43.32 \pm 0.23 $ & $0.095  \pm 0.011$ & $13.02 \pm 0.27$  \\ 
 \hline
541  &  $62.20 \pm 1.5 $ & $0.135  \pm 0.015$ & $15.52 \pm 0.07 $  \\ 
$1800^{(a)}$ &  $72.20 \pm 2.7 $ & $0.140  \pm 0.069$ & $16.72 \pm 0.44 $  \\ 
$1800^{(b)}$ &  $80.03 \pm 2.24 $ & $0.15    $ & $16.98 \pm 0.25 $  \\  
\hline
   \end{tabular}
   \end{table}
   \end{center}

Our determination of the parameters is described below. 
In  Table \ref{our_results} we collect our results. It is remarkable that 
the obtained $\chi^2$ values are very small. 

\begin{center}
   \begin{table}
   \caption{ \label{our_results} Forward scattering parameters obtained in our analysis. }
   \vspace{.5 cm}
   \begin{tabular}{|c|c|c|c|c|c|c|}
   \hline
$\sqrt{s}$ (GeV)&$\sigma ({\rm mb})$& $\rho$ & $B_I({\rm GeV}^{-2})$ & $B_R({\rm GeV}^{-2})$& $\beta=B_R/B_I$ &$\chi^2$ \\
 \hline
$ 19.4 $  & $40.379 \pm 0.069$& $0.019$ ({\rm fixed})       & $14.539 \pm 0.262 $& $ B_I ~ ,~ 2 B_I $ & 1 ~ ,~ 2      & $1.299 $ \\ 
\hline
$ 23.5 $  & $39.821 \pm 1.479$& $0.0186\pm 0.0137$ & $14.912 \pm 9.246 $& $ 35.220\pm 177.57$ & $2.36$  & $0.2952$ \\ 
  \hline
$ 30.7 $  & $40.024 \pm 0.047   $& $0.027 ({\rm fixed})    $ & $ 11.784 \pm 0.239  $& $ 
 B_I ~ , ~ 2 B_I   $     & $1 ~, ~2 $  & $ 0.5361  $ \\ 
  \hline
$ 44.7 $  & $41.839 \pm 0.291$& $0.0543\pm 0.0037$ & $12.976 \pm 0.631 $& $ 16.132\pm 15.509 $ & $1.243$ & $0.6110$ \\ 
  \hline
$ 52.8 $  & $42.576 \pm 0.820$& $0.0799\pm 0.0086$ & $13.414 \pm 1.847 $& $ 14.113\pm 33.684 $ & $1.052$& $0.1138$ \\ 
  \hline
$ 62.5 $  & $43.298 \pm 0.159$& $0.0867\pm 0.0034$ & $13.299 \pm 0.358 $& $ 13.900\pm 10.007 $ & $1.045$& $0.5389$ \\ 
  \hline
  \end{tabular}
  \end{table}
  \end{center}

\subsection{pp scattering at $\sqrt{s}$ = 19.4 GeV} 
\begin{figure}
 \caption {Data from Kuznetsov (61 points) \cite{kuznetsov} and 
Schiz (12 points)\cite{schiz} are fitted  to determine parameters
at 19.4 GeV. We fix $\rho=0.019$ (taken from Table \ref{basic}) 
because fittings including Kuznetsov's measurements  lead to negative 
value of $\rho$. } 
\label{data_19_4}
\includegraphics[height=10cm]{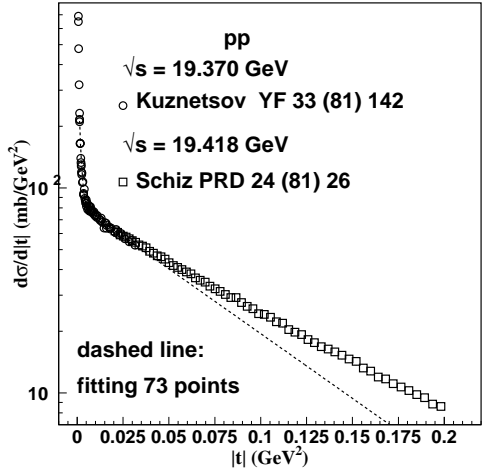}
 \end{figure}

Considering only Kuznetsov \cite{kuznetsov} and Schiz \cite{schiz} measurements, 
we have a total of 69+134 points at 19.4 GeV.  We have fittted the set of the 
first 61 points  
from Kuznetsov   plus the 12 first ones from Schiz, covering the range
$$ 0.00066 \leq |t| \leq 0.0395 ~. $$
According to the information in the Durham Data Basis
about this experiment, it is known that fittings that include the points 
of Kuznetsov lead no negative values of the parameter $\rho$.  
We then 
fix the value  $\rho=0.019$ (taken from Table \ref{basic}) and leave free the other parameters. The 
results are given in Table  \ref{our_results} . 
The data and the solution of fitting are shown if Fig. \ref{data_19_4} .
It is remarkable that the choices $B_R=B_I$  and $B_R=2 ~ B_I$ lead to the 
same $\chi^2$.  

\subsection{pp scattering at $\sqrt{s}$ = 23.542 GeV} 
At 23.542 GeV there are 31  experimental points \cite{amos}.
In order to obtain smaller values of $\chi^2$, the fitting was made using 
the first 17 points, with $|t|$ in the interval 
  $$    0.00037   \leq |t| \leq   0.00395   ~ ,   $$ 
leading to the results given in Table  \ref{our_results}. 
 The central value obtained for the ratio $\beta$ is  
$$ \beta= B_R / B_I = 2.36   ~ . $$  
The same value $\chi^2=0.2952$ is obtained for any $\beta = B_R/B_I$ in the 
interval  from  1.22 to 4.91 .

\begin{figure}
 \caption { Data at 23.542 GeV . The first 17 points are fitted 
 with expressions of Coulomb interference and exponential forms
 for the amplitudes. The graph in the RHS shows the regions of 
variation of $\rho$ and $\beta=B_R/B_I$ that correspond to small 
 values of $\chi^2$ while the quantities $\sigma$ and $B_I$ are free. } 
\label{data_23}
\includegraphics[height=10.0cm]{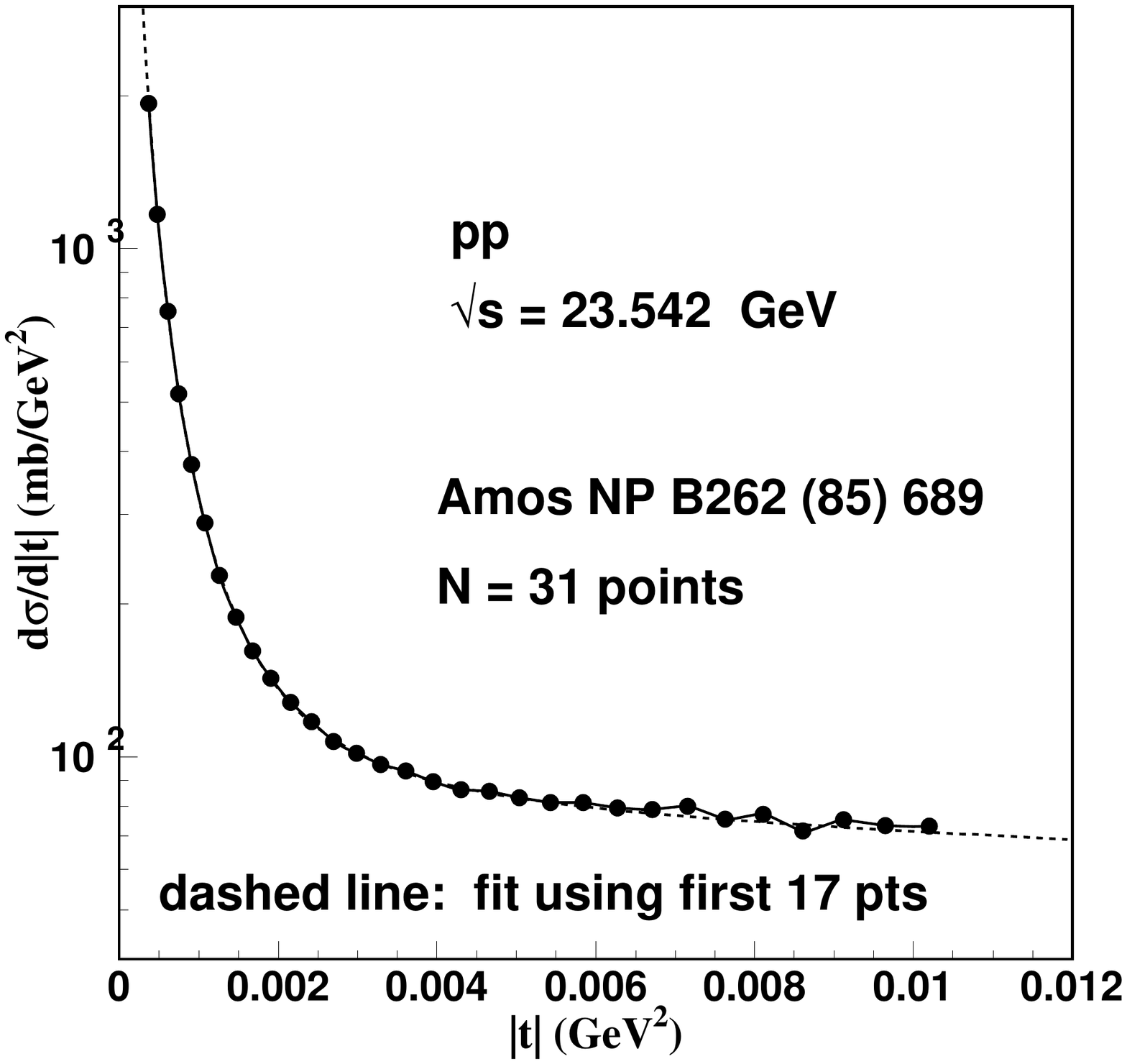}
\includegraphics[height=10.0cm]{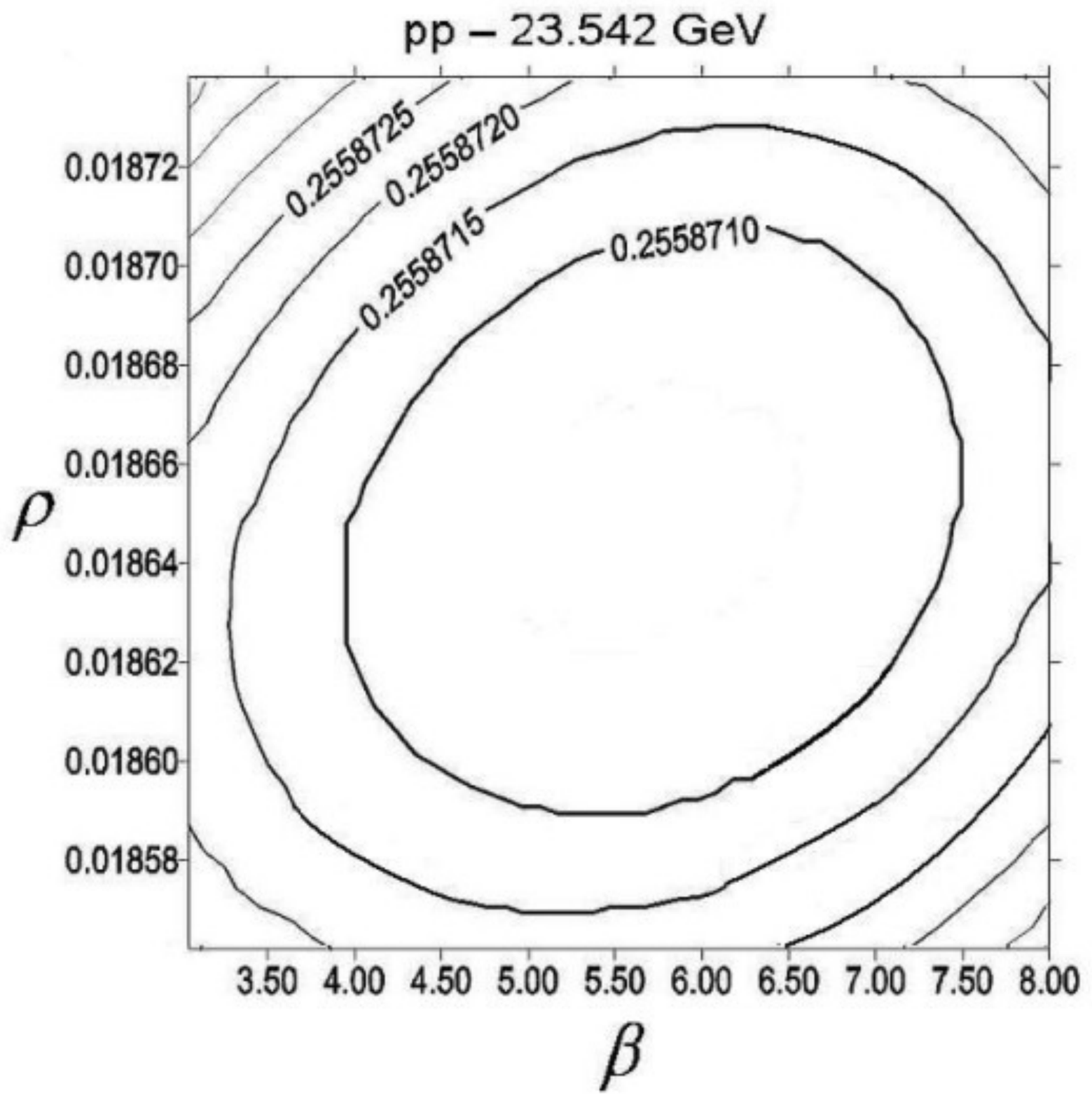}
 \end{figure}

The plot of the 31 points together with the line 
  obtained with fitting of 17 points  are shown in Fig \ref{data_23}. 
Correlations between parameters $\rho$ and $\beta$ are shown in the RHS, 
with level curves of $\chi^2$. This plot is made fixing the values of 
$\rho$ and $\beta$, so that only 2 parameters are free 
(this explains the relation $0.2952=0.25587\times(17-2)/(17-4)$ in the 
values of $\chi^2$).  The two different algorithms lead to distinct 
ranges in $\beta$, indicating that the data have poor definition for
this quantity.  
 
\subsection{pp scattering at $\sqrt{s}$ = 30.632 GeV} 

\begin{figure}
 \caption { Data at 30.632  GeV . The 32 points are fitted 
 with expressions of Coulomb interference and exponential forms
 for the amplitudes. } 
\label{data_30}
\includegraphics[height=10cm]{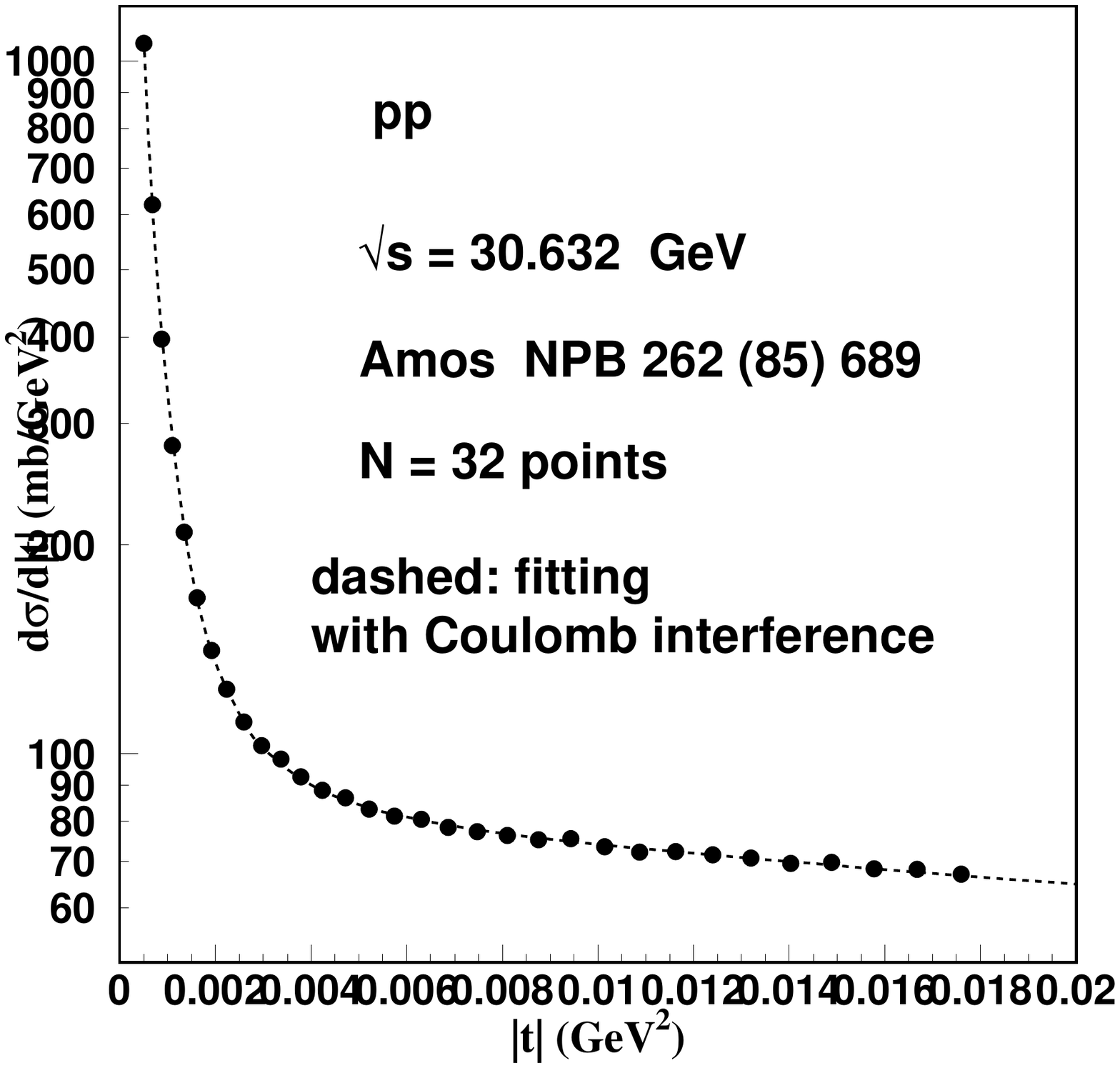}
 \end{figure}

Although the data (32 points) of pp scattering  at 30.632 GeV \cite{amos} 
look regular , 
we have difficulties to find values for the parameters, although $\chi^2$ 
 comes out small.  
The value 0.042 given for $\rho$  in Table \ref{basic} does not 
seem to be realistic, as our procedure leads to smaller values. 
With the value $\rho=0.027$ fixed, the  ratio $\beta=B_R/B_I$ 
can vary in a large interval, keeping the same $\chi^2=0.5361$.
The data, fitted with Coulomb interference expressions and exponential 
amplitudes,  are shown in Fig. \ref{data_30} and the parameters are 
given in Table \ref{our_results}. 
  
\subsection{pp scattering at $\sqrt{s}$ = 44.699 GeV} 
\begin{figure}
 \caption {Fitting of the 40 points with lowest  $|t|$ at 44.699 GeV .   } 
\label{forward_44}
\includegraphics[height=10cm]{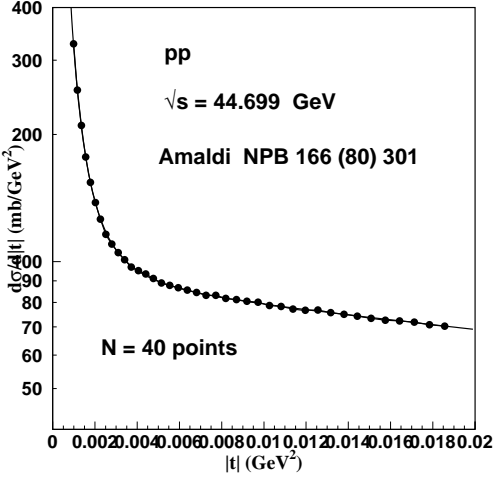}
 \end{figure}

The 230 data points of the experiment at $\sqrt{s} = 44.699$  GeV 
extending up to $|t|\approx 7 ~ {\rm GeV}^2$  are  presented in the 
report by  Amaldi and Schubert \cite{amaldi}.  
The forward part, with 40 points,  up to 
$|t| \approx 0.02 ~ {\rm GeV}^2$,  is very well fitted by the Coulomb 
interference formula, as shown in Fig. \ref{forward_44}, with 
$\chi^2=0.6110$. The parameters are given in Table  \ref{our_results} .

\begin{figure}
 \caption {The extended dashed line representing the fitting at very 
low $|t|$ is    plotted together with measurements at higher $ |t| $, 
at $\sqrt{s}=44.699$ GeV. The dotted line shows a fitting of the whole 
set of points \cite{pereira_ferreira}.} 
\label{intermediate_44}
\includegraphics[height=9cm]{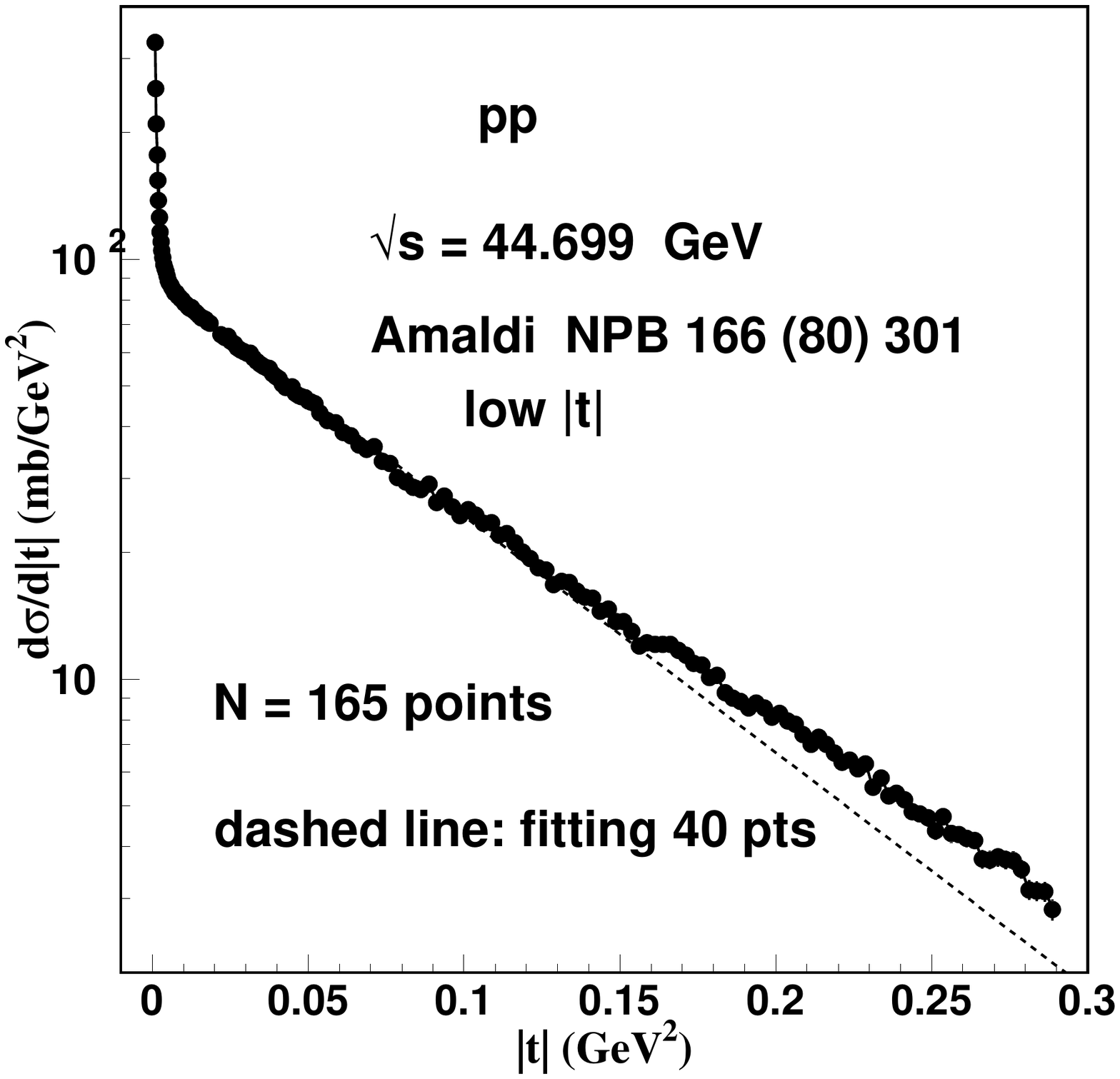}
 \includegraphics[height=9cm]{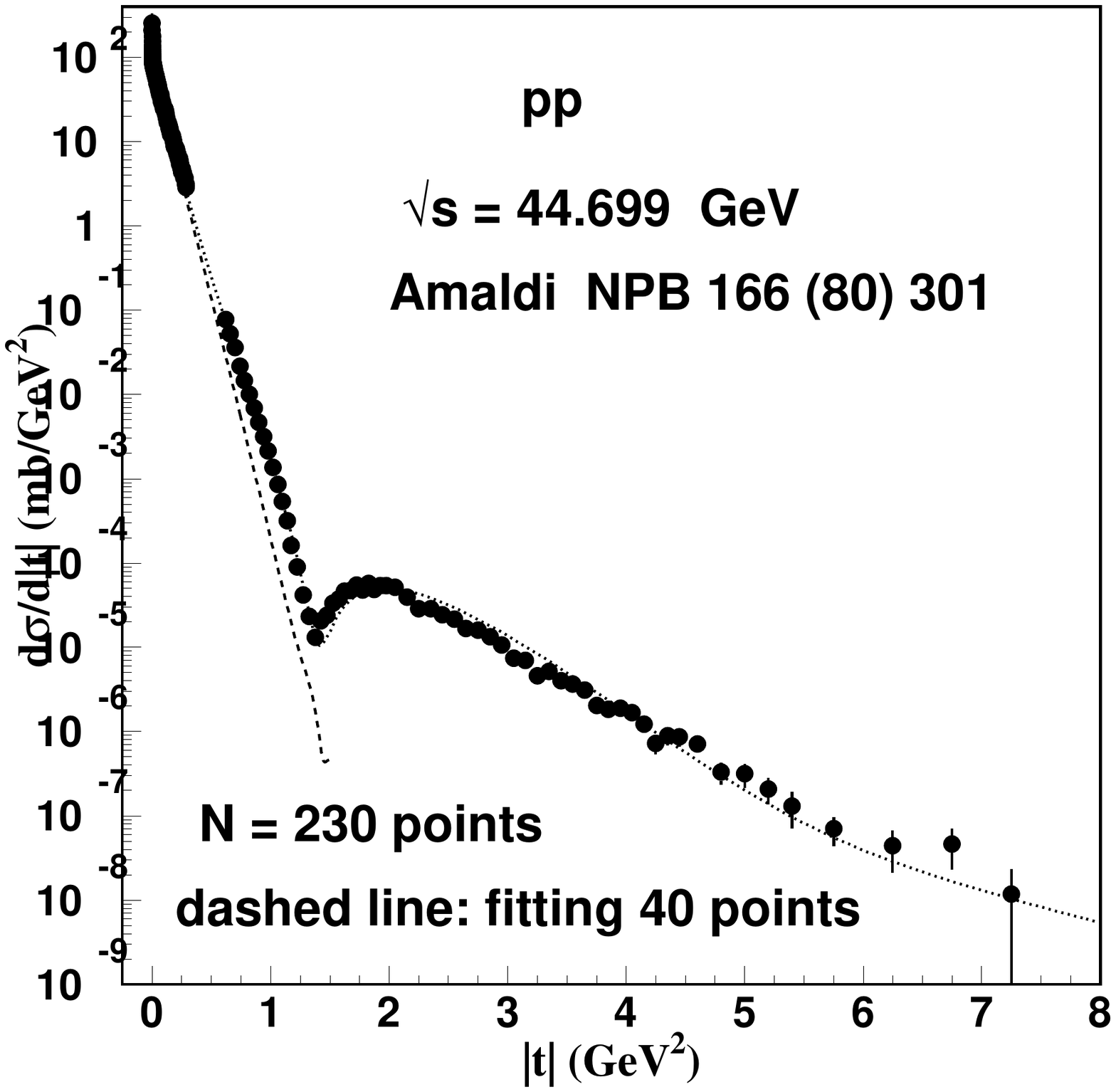}
 \end{figure}

\begin{figure}
 \caption {Ranges of values of parameters at $\sqrt{s}=44.699$ GeV 
that lead to small values of
    $\chi^2$. The searches are made fixing the two values in the axes 
of the plots, while the  other two  parameters are free. } 
\label{bolhas_44}
\includegraphics[height=10.0cm]{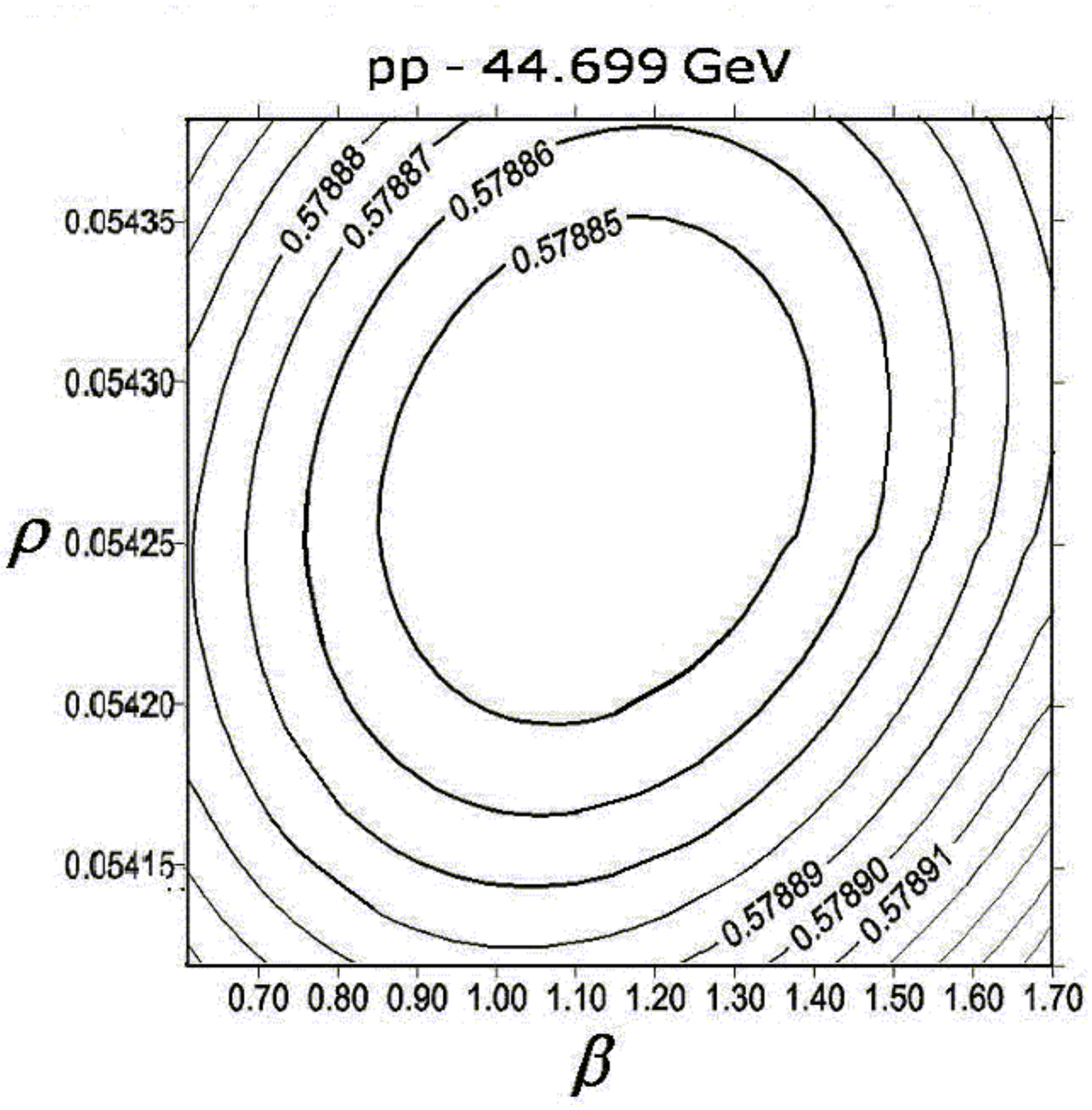}
 \includegraphics[height=10.0cm]{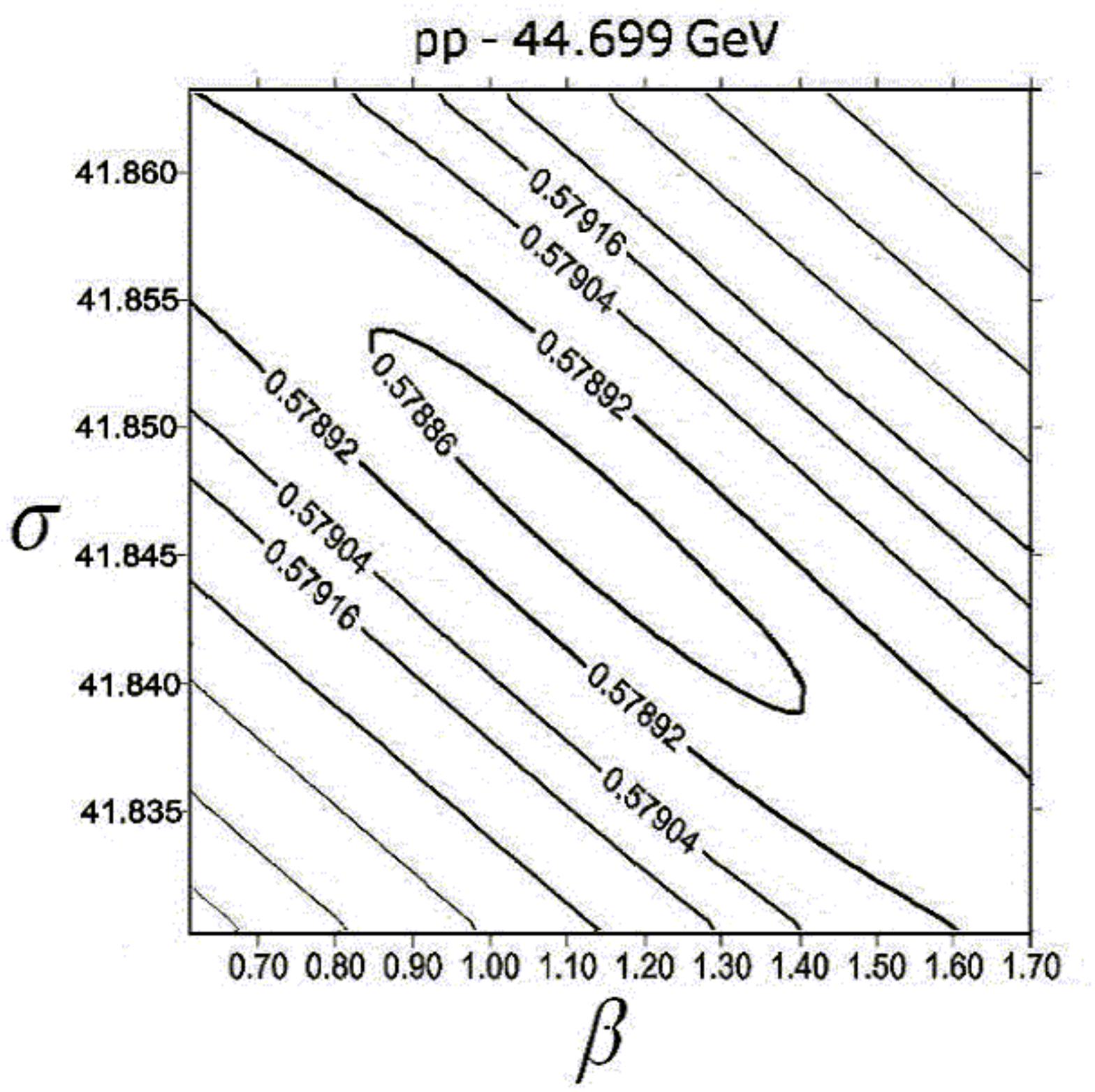}
 \end{figure}

 Any value of the ratio $ \beta={B_R}/{B_I} $ in the range  
$$ 0.601 ~ \leq  ~ \frac{B_R}{B_I} ~  \leq  ~ 1.667    $$ 
leads to the same value 0.6110 for $\chi^2$.  
In Fig. \ref{intermediate_44} we show the solution for low $|t|$ together with higher $ |t| $ data, exhibiting the peculiar behaviour of the 
amplitudes deviating from the simple exponential dependence. 
In the plot with all 230 points , the dotted line shows the fitting 
obtained with the parametrization used in a previous work 
\cite{pereira_ferreira}.  The parameters obtained in this case are 
$\sigma=42.10$ mb, $\rho=0.094$, $B_I=12.22 ~ {\rm{GeV}}^{-2}$,  and     
$B_R=24.43 ~ {\rm{GeV}}^{-2}$. 
It is interesting that the description of the whole $|t|$  range 
leads to  a definite indication for $B_R \approx 2~B_I$. 
Correlations between parameters are shown in Fig \ref{bolhas_44}, 
with level curves of $\chi ^2$ . The two plots show respectively 
the correlations between $\rho$ and $\beta$ and between $\sigma$ 
and $\beta$. of 

With fixed $B_R/B_I=2$ we obtain $\chi^2=0.6111$ , with $\sigma=41.819\pm 0.089$ , 
$\rho=0.0543 \pm 0.0029$ , $B_I=12.943 \pm 0.406$ . 

%  \clearpage
  
\subsection{pp scattering at $\sqrt{s}$ = 52.806 GeV} 
\begin{figure}
 \caption {Data of pp forward scattering at 52.8 GeV fitted with Coulomb 
  interference formulae.} 
\label{data_52_8}
\includegraphics[height=10.0cm]{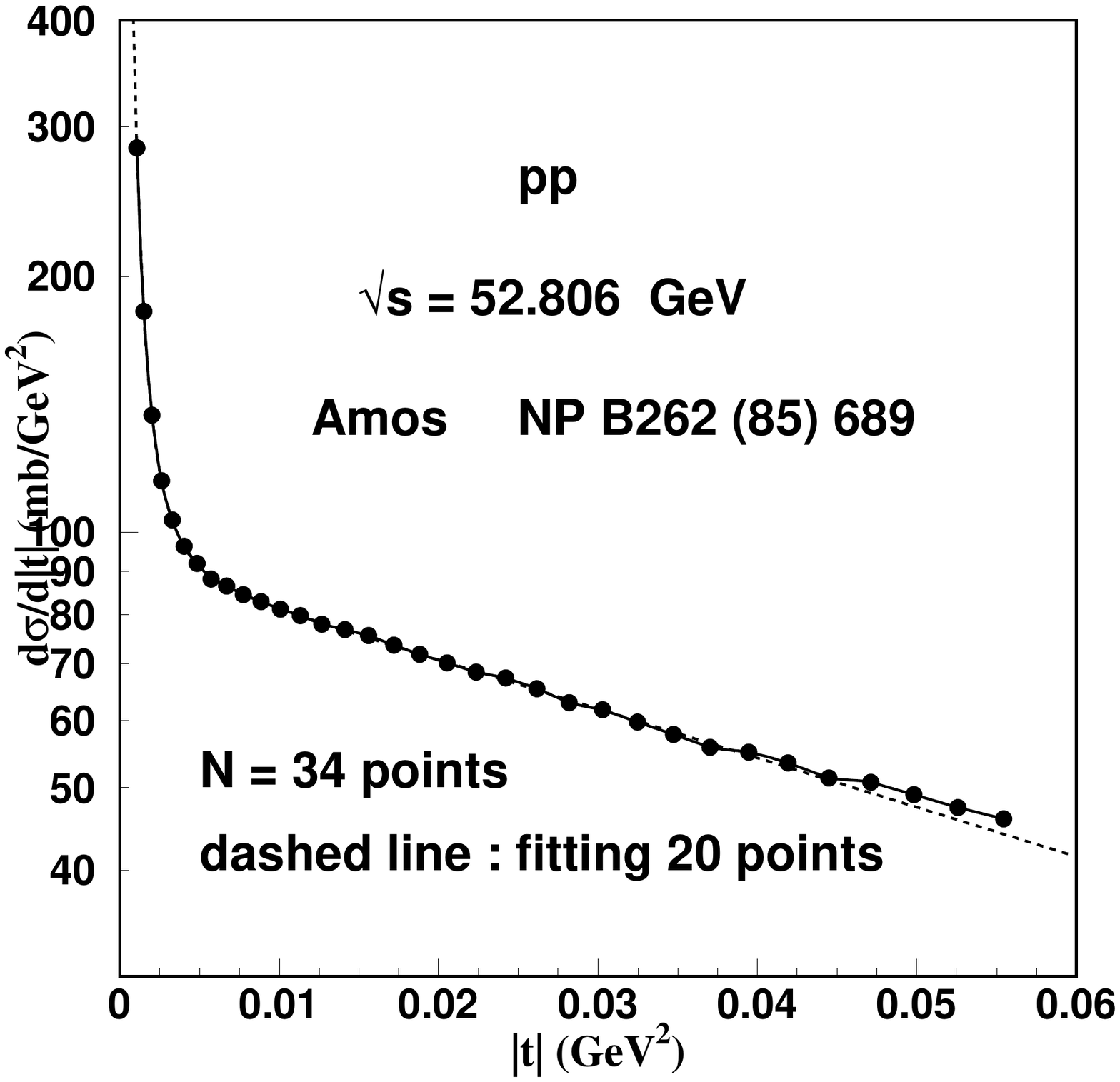}
 \end{figure}

\begin{figure}
 \caption {Correlations between pairs of parameters in pp scattering at 
  52.8 GeV. } 
\label{bolhas_52}
\includegraphics[height=10.0cm]{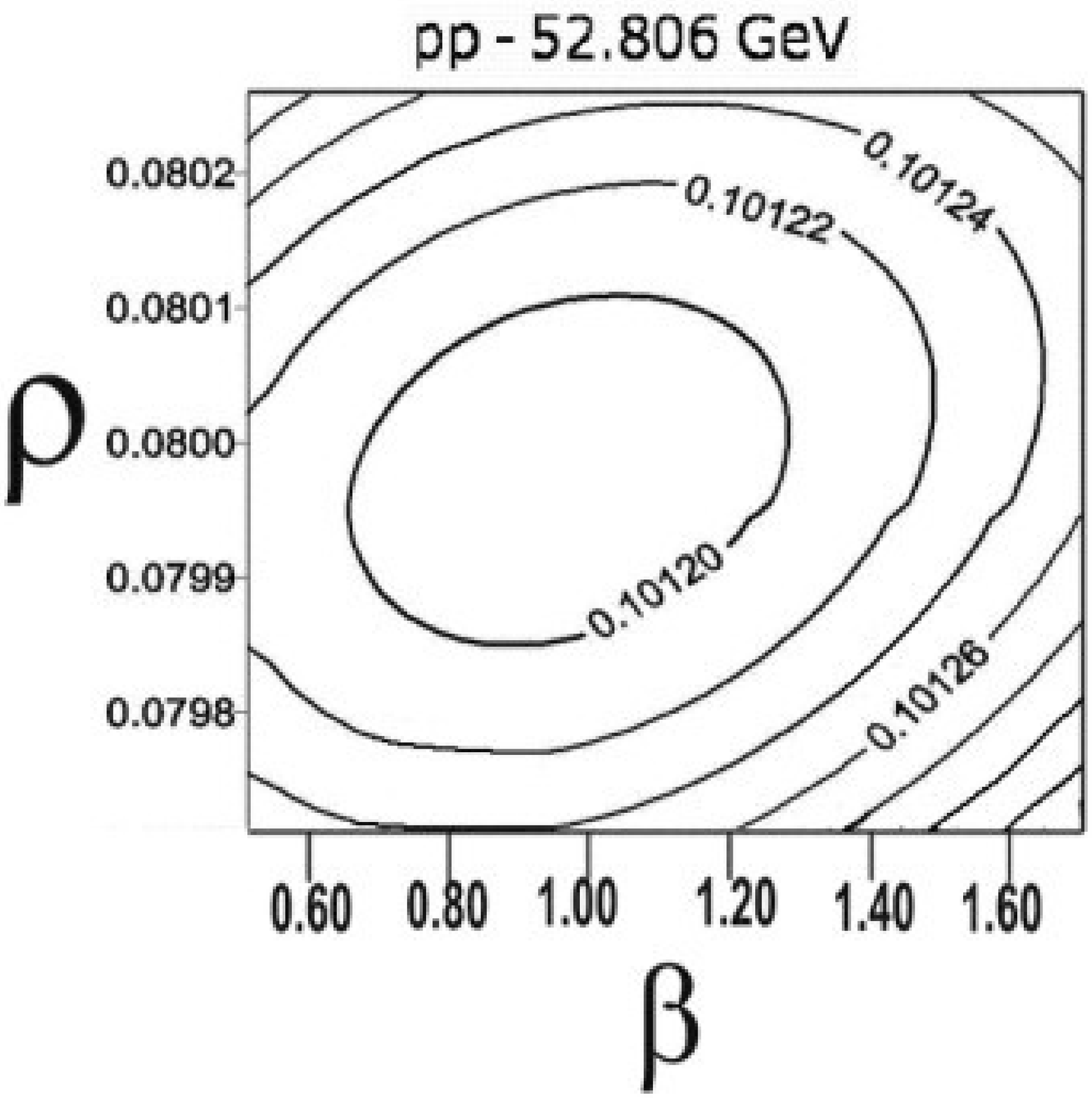}
 \includegraphics[height=10.0cm]{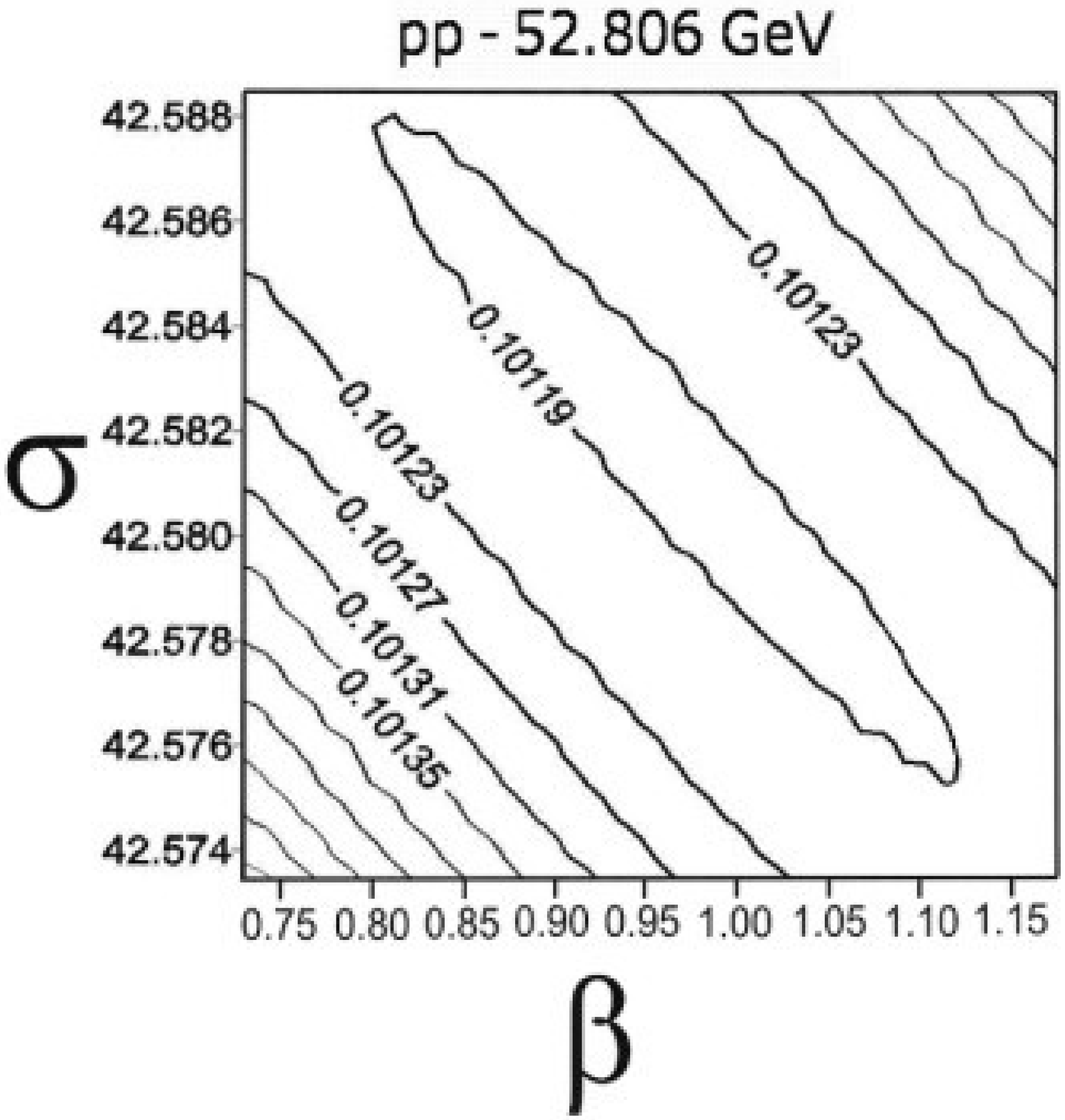}
 \end{figure}

Fig. \ref{data_52_8} shows the forward data at 
$\sqrt{s}=52.806$ GeV (34 points) \cite{amos} and our 
fitting with Coulomb interference expressions using the first 20 points.
 Correlations between parameters are shown in Fig. \ref{bolhas_52} , 
with level curves of $\chi ^2$ . The two plots are built 
fixing the two parameters in the axes while the other two parameters 
are found by fitting. 
The parameters obtained with 20 points  in the interval 
     $$ 0.00107 \leq  |t| \leq 0.02235   $$
 are  given in Table  \ref{our_results} ~ . 
The  ratio $ \beta={B_R}/{B_I} $ with any value in the range  
   $$ 0.702 \leq  B_R / B_I \leq 1.214  ~  $$ 
 leads to the same value 0.1138  for $\chi^2$. 

With fixed $B_R/B_I=2$ we obtain $\chi^2=0.1140$ , with $\sigma=42.543 \pm 0.057$ , 
$\rho=0.0801 \pm 0.0047$ , $B_I=13.347 \pm 0.214$ .

 \subsection{pp at $\sqrt{s}$ = 62.5 GeV} 
\begin{figure}
 \caption { Data (138 points) of pp scattering at 
$\sqrt{s}=62.5$ GeV  \cite{amos} shown together with
our fitting of the first 40 points of the set with Coulomb interference expressions. } 
\label{data_62_5}
\includegraphics[height=9.0cm]{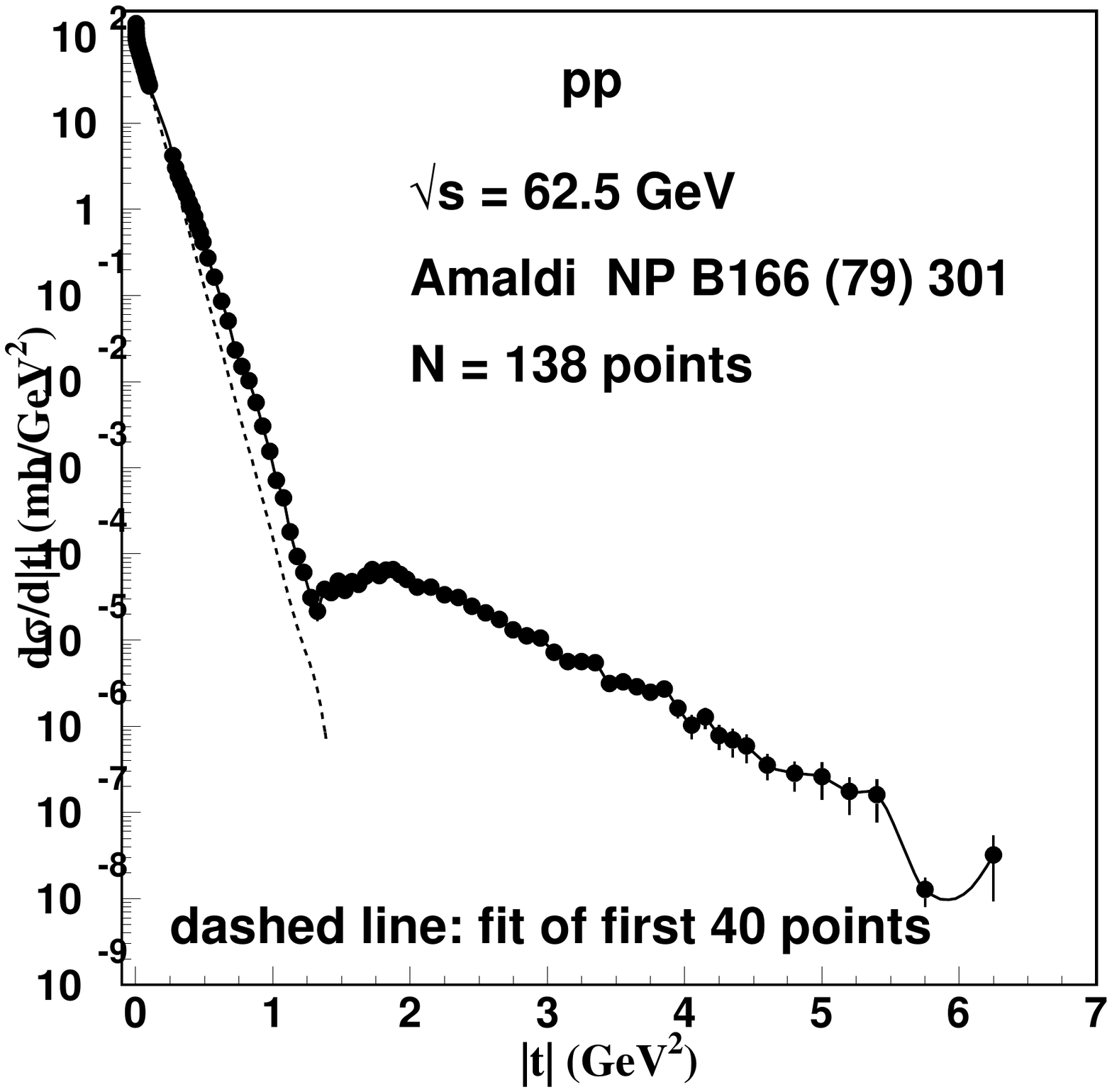}
\includegraphics[height=9.0cm]{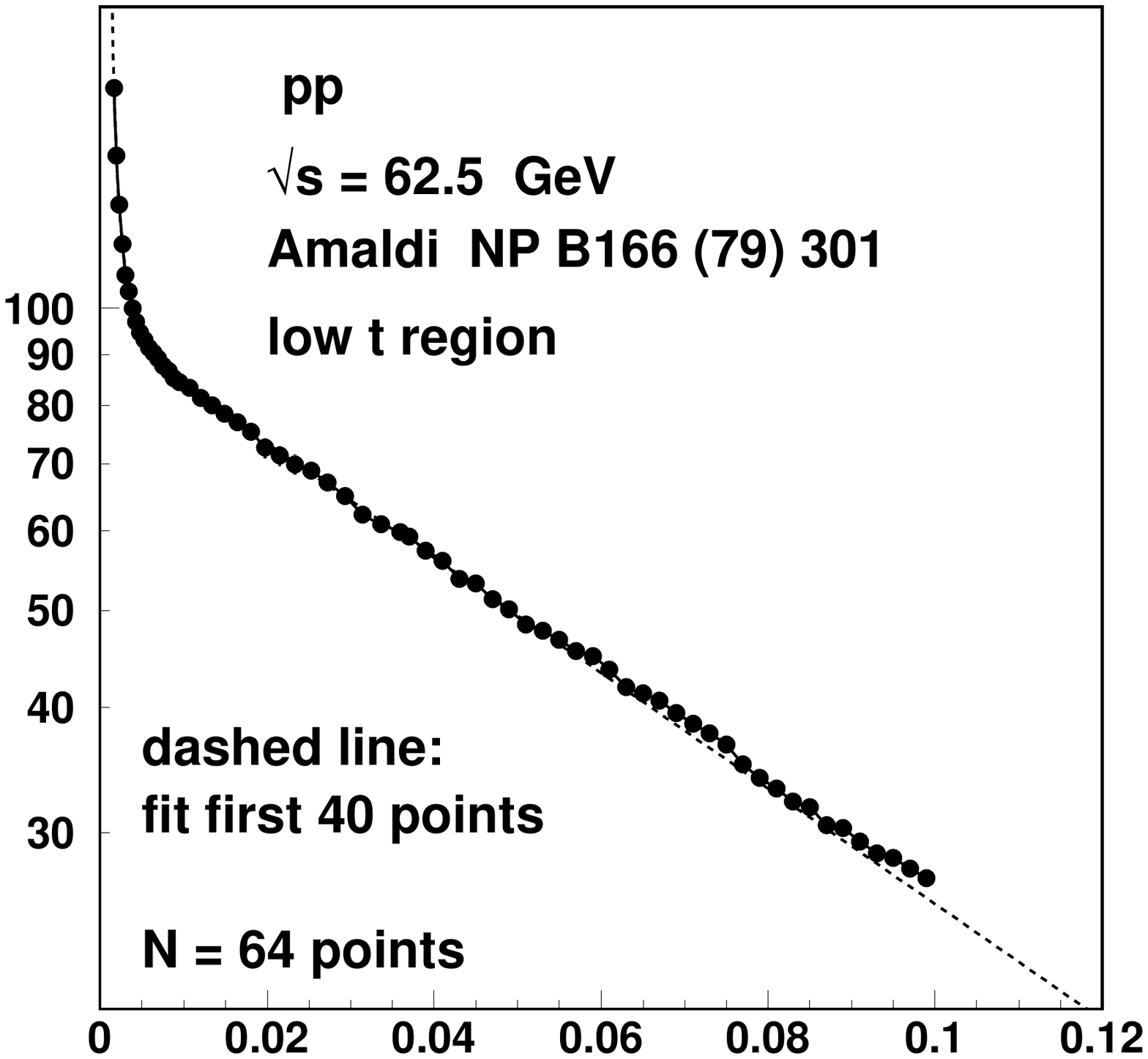}
 \end{figure}

\begin{figure}
 \caption {Correlations between pairs of parameters 
( $ \rho ~ - ~ \beta $ and $\rho ~ - ~ \sigma$   
in pp scattering at 
  62.5 GeV. }   
\label{bolhas_62}
\includegraphics[height=10.0cm]{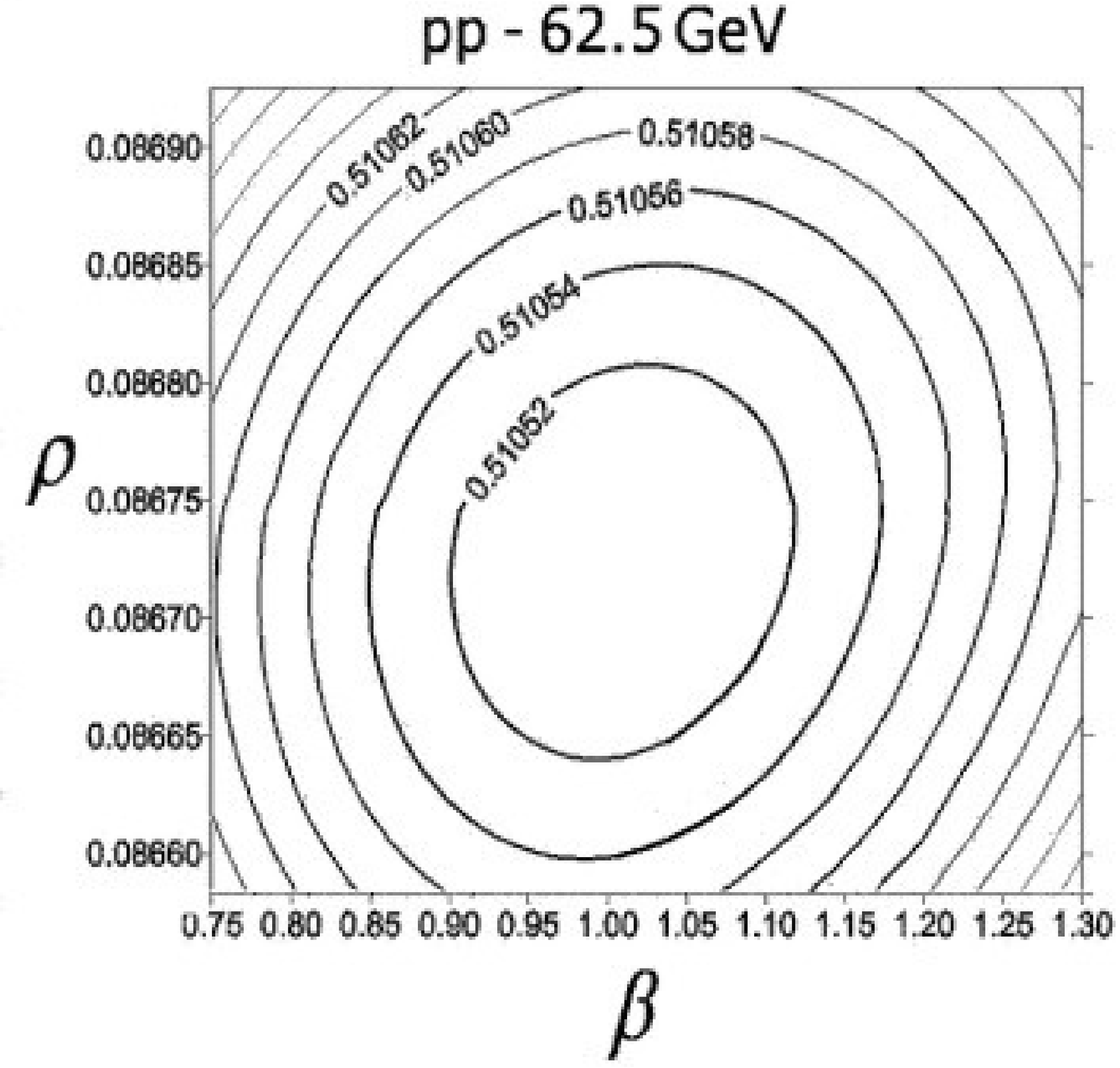}
\includegraphics[height=10.0cm]{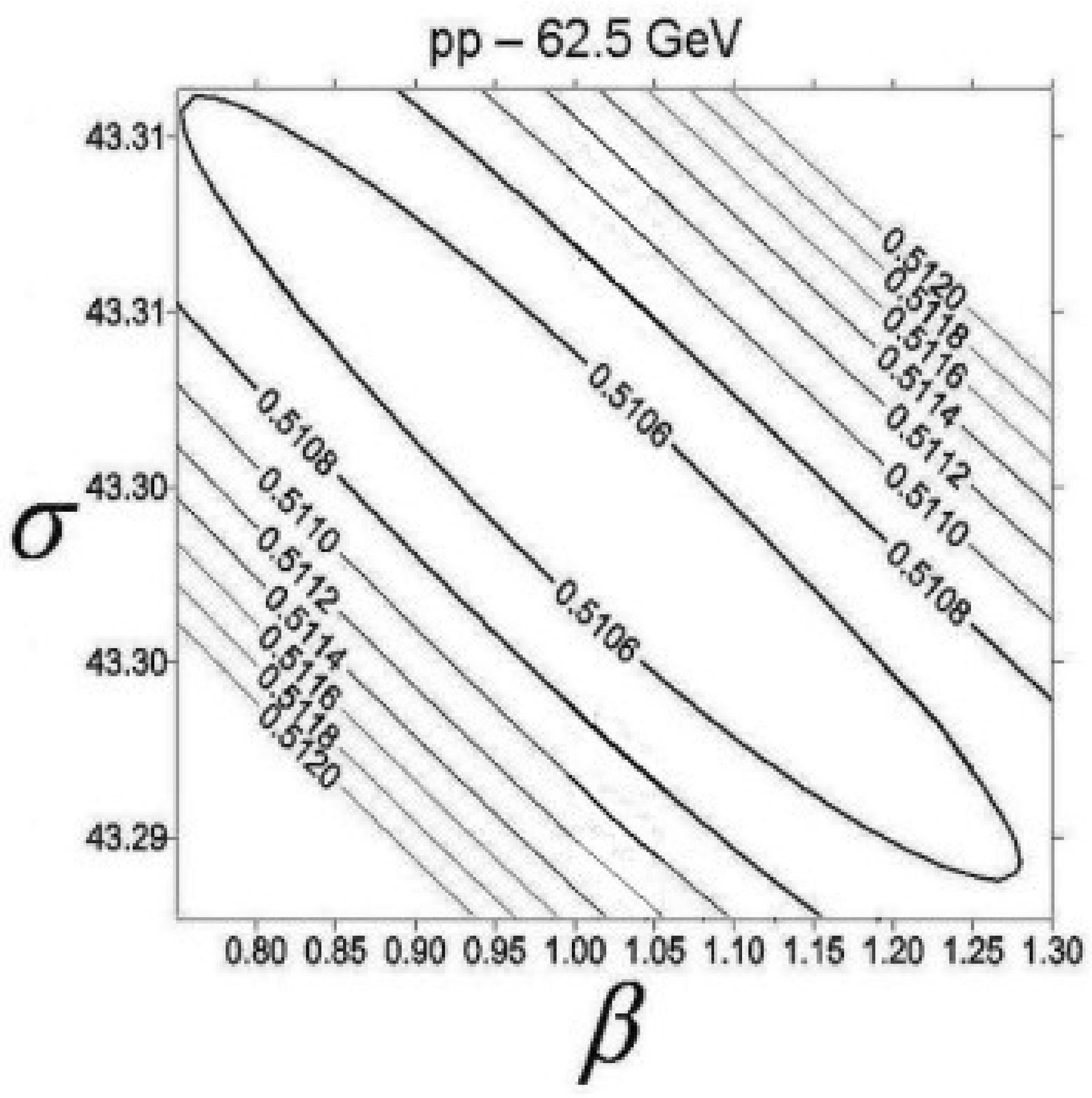}
 \end{figure}

 Fig. \ref{data_62_5} shows the data (138 points) \cite{amos} and result 
of our fitting with Coulomb interference expressions using the first 
40 points of the set.
Correlations between parameters are shown in Fig. \ref{bolhas_62} , 
with level curves of $\chi ^2$ .
The parameter values obtained in fitting with 40 first points, 
in the $|t|$ interval 
      $$ 0.00167 \leq  |t| \leq 0.051  $$
are given in Table \ref{our_results}. 
 The same $\chi^2 = 0.5389$  is obtained with ratio $\beta$ with any 
value in the interval 
$$ 0.78 \leq  B_R / B_I \leq 1.25  ~ . $$ 

With fixed $B_R/B_I=2$ we obtain $\chi^2=0.5400$ , with $\sigma=43.264 \pm 0.025$  mb  , 
$\rho=0.0869 \pm 0.0017$ , $B_I=13.230 \pm 0.077 {\rm GeV}^{-2}$ . 
 
%  \clearpage 

\subsection{Analysis of the event rate dN/dt at $\sqrt{s}$ = 541 GeV}

The lowest $|t|$ values reached in  measurements of p$\bar {\rm p}$ 
elastic scattering in the neighborhood of 540 GeV are reported with event 
rates \cite{augier} only. The cross-section values have not been 
determined otherwise in this low $|t|$ range, according  to Durham HEP data 
basis.  In the present work we use the Coulomb interference as a 
tool to find the normalization factor connecting event rate  and 
differential cross-section. We find a very clear and precise connection,
which is very important, as the event rate $dN/dt$ has been measured 
with homogeneous  accuracy, with many (99) points  in  a range of low  
$|t|$ values.   Our simple procedure is to fit the $dN/dt$ data 
with the expression for the Coulomb interference region, with 
and arbitrary multiplying normalization factor, and the four 
parameters that describe forward p$\bar {\rm p}$ scattering.   

\begin{figure}
 \caption { Event rate \cite{augier} and differential cross section at 541 GeV . }
\label{sig541}
\includegraphics[height=9cm]{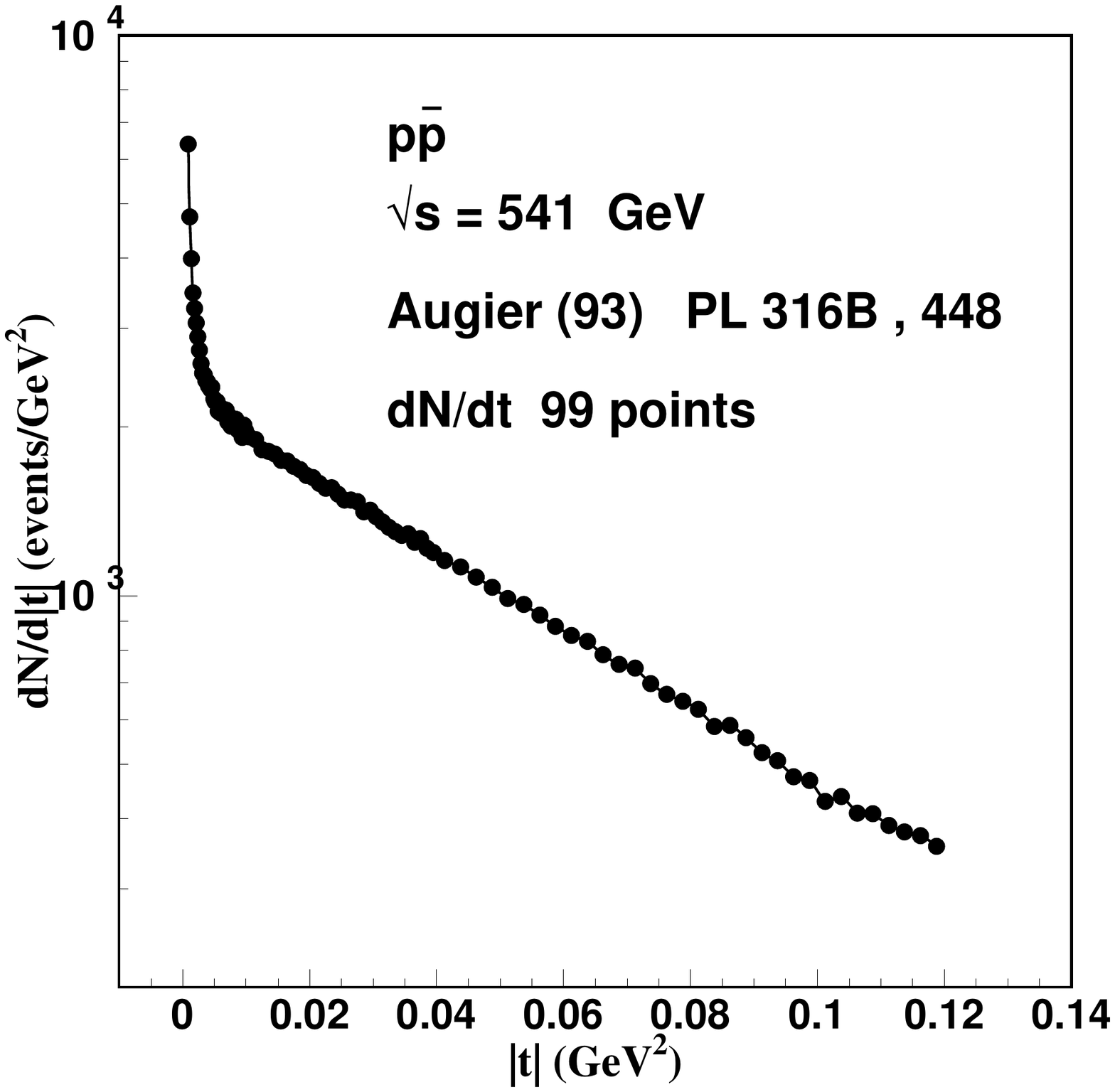}
 \includegraphics[height=9cm]{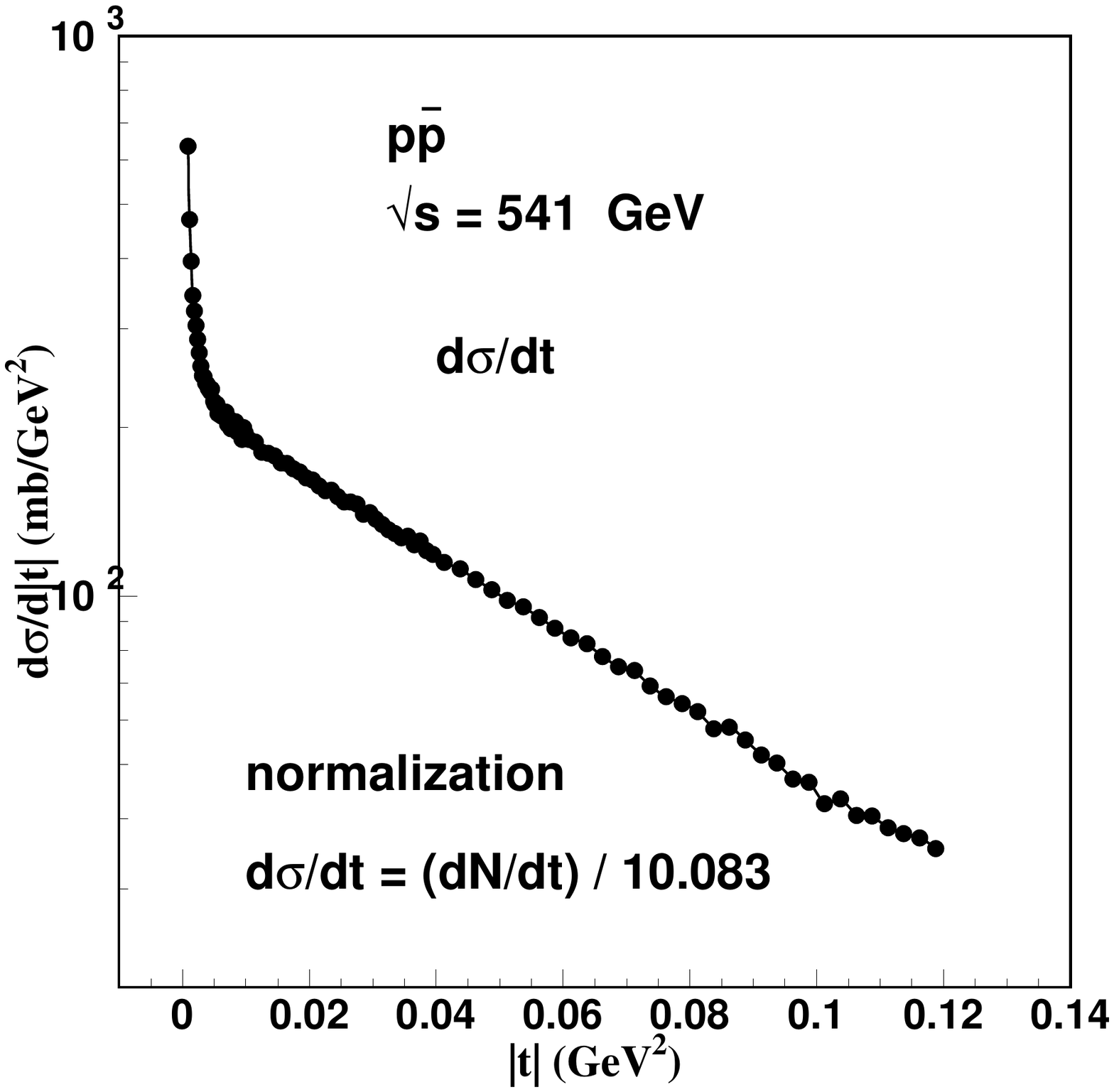}
 \end{figure}

\begin{figure}
 \caption { The normalized event rate data at 
541 GeV re in good agreement with the $d\sigma/dt $ data 
\cite{arnison,bozzo-I, bozzo-III} at the same energy and higher 
$|t|$ values. } 
\label{compare}
\includegraphics[height=9cm]{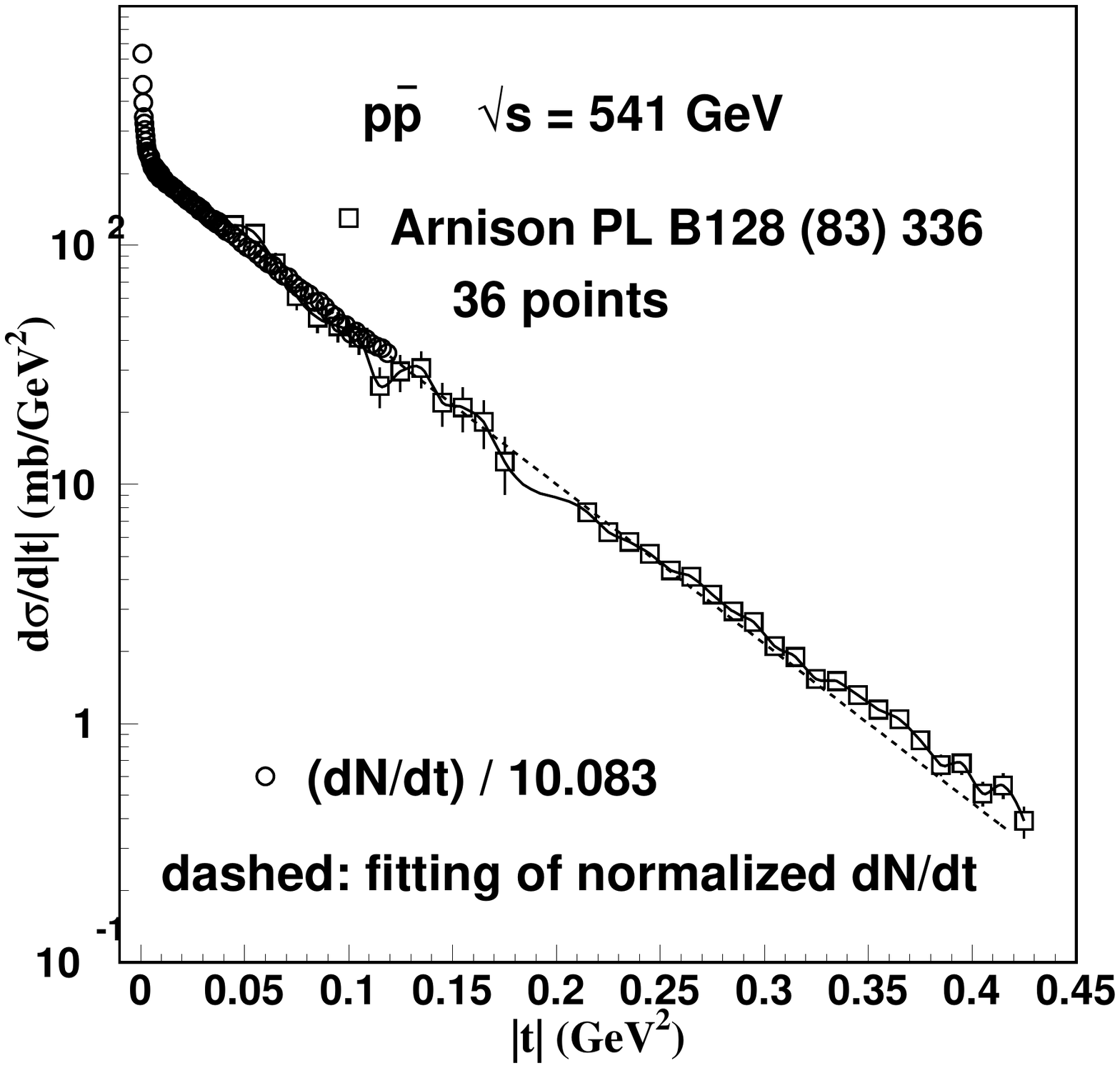}
 \includegraphics[height=9cm]{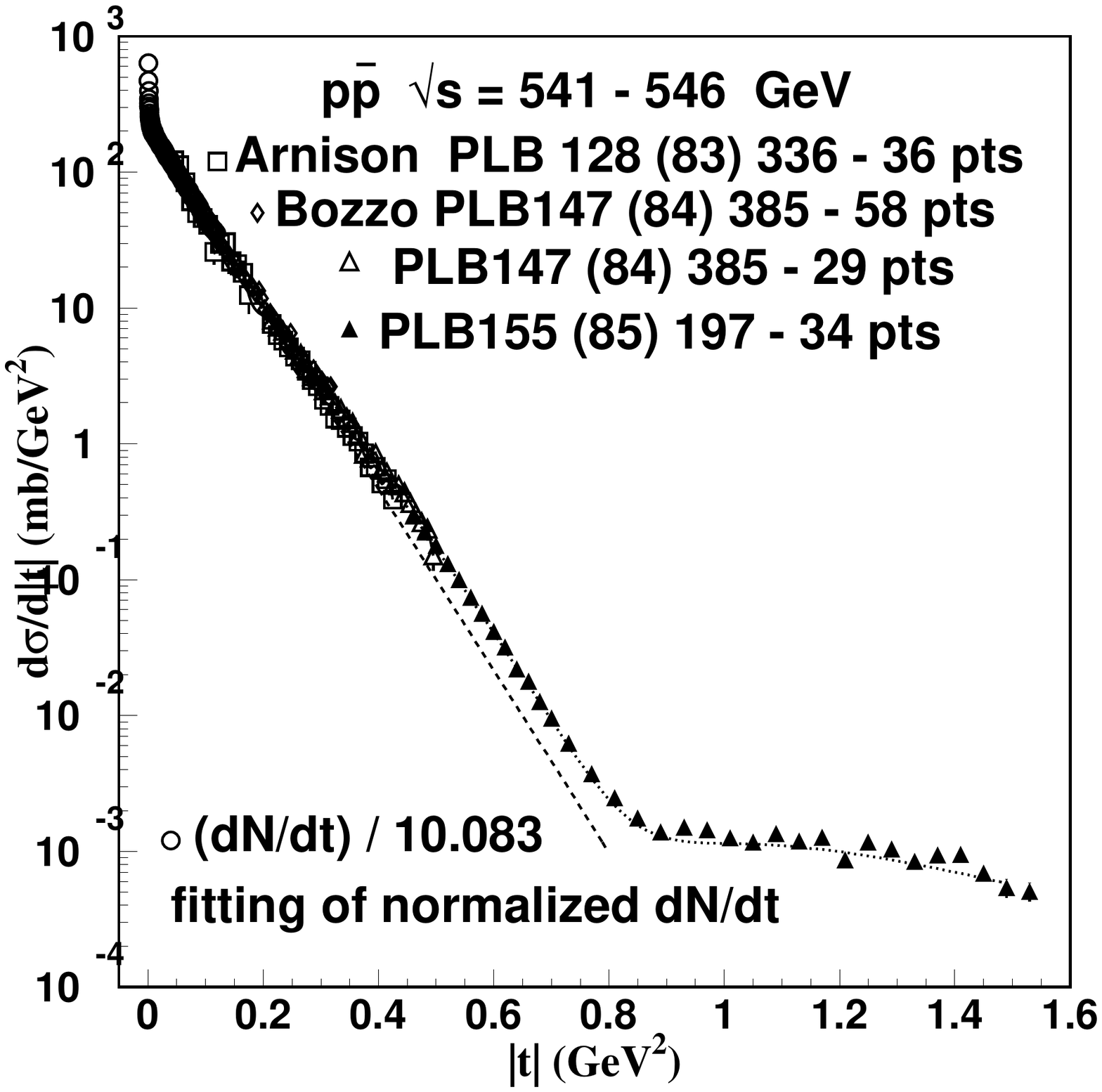}
 \end{figure}

Fig. \ref{sig541} shows the 99 points of the $dN/dt$ measurements 
\cite{augier},  of p$\bar {\rm p} $ elastic scattering at 541 GeV 
and the values of the differential cross section after 
we determine the normalization by Coulomb interference. 
We have found the normalization factor 
\begin{equation}
\label{normalization}
  \frac{d\sigma}{dt}=\frac{d\sigma}{dt} \times \frac {1}{10.083 \pm 0.135}   
\end{equation}
In Fig. \ref{compare} we show the good agreement   
of $d\sigma/dt$ at low $|t|$ , obtained from  dN/dt by 
adjustment of the coulomb interference, with the 
data of G. Arnison et al \cite{arnison} 
 in a $|t|$ range that partially superposes  with the 
$dN/dt$ event rate data.  In the RHS of the same figure 
 we include also the data of Bozzo et al. 
\cite{bozzo-I, bozzo-III}, including high $|t|$ values. 

As a test of consistency of this method of connection between event rate
and absolute cross section we compare values of $d\sigma/dt$ at the Cern/ISR 
energies, multiplied by an arbitraty normalization factor,  with the 
Coulomb interference amplitudes. We find that this normalization factor is   
actually equal to one in all investigated cases.

\begin{figure}
 \caption { Looking closely at the normalized cross section 
obtained with factor 10.083 derived purely from Coulomb 
interference formulae, compared to Bozzo et al. data 
\cite{bozzo-I, bozzo-III}, 
we observe that there remains a displacement of a few percent. 
A more precise  matching is obtained with a factor (dN/dt)/10.6 , 
as shown by the dashed line in the figure.  } 
\label{normal_bozzo}
\includegraphics[height=10cm]{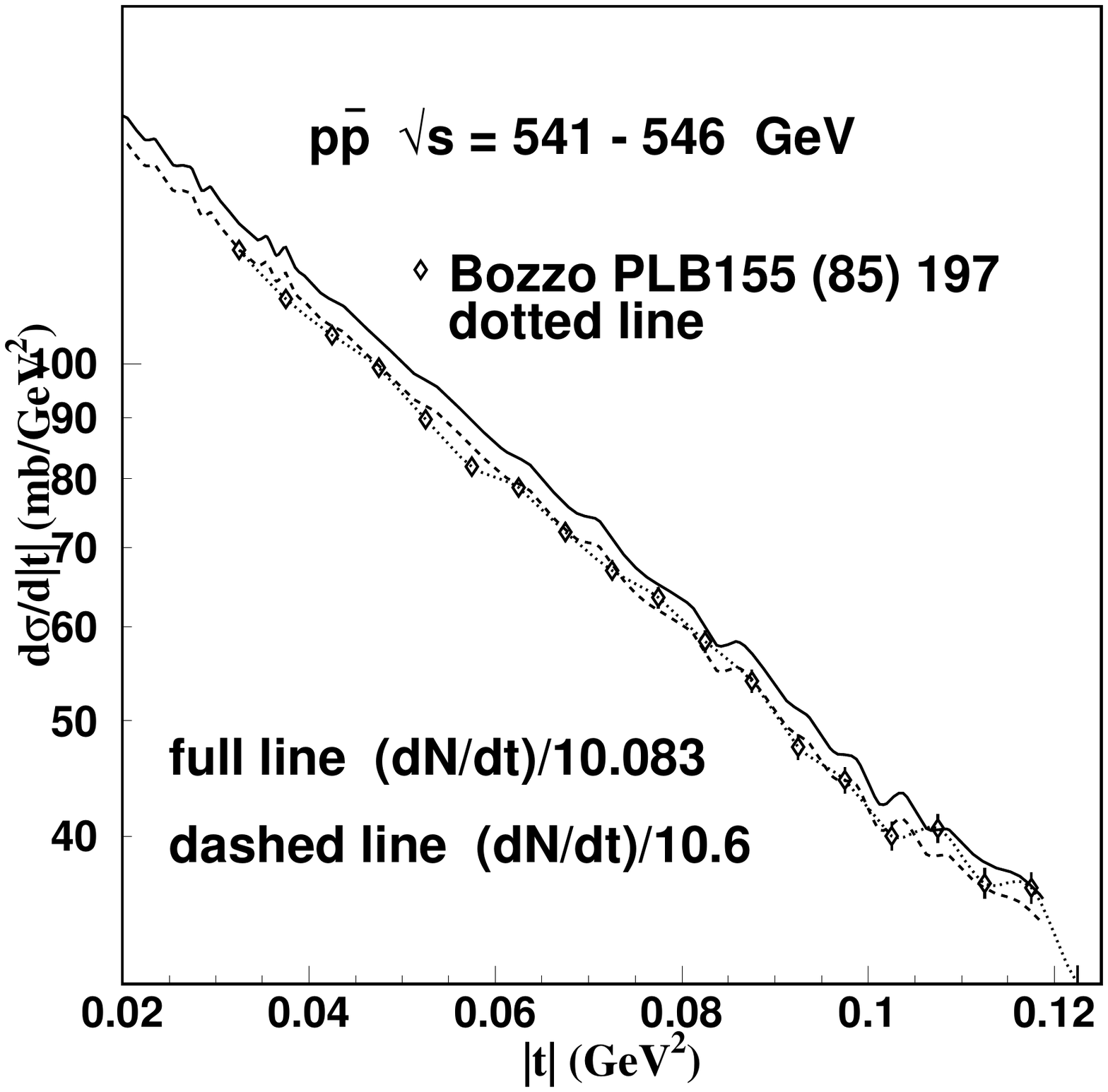}
 \end{figure}

\begin{figure}
 \caption { The measurements of Bernard et al. \cite {bernard} 
and Abe et al. \cite {abe} are compatible with each other, but 
do not match the other measurements and 
our conversion from the event rate, requiring an arbitrary 
matching factor 11.0 . } 
\label{bernard_abe}
\includegraphics[height=10cm]{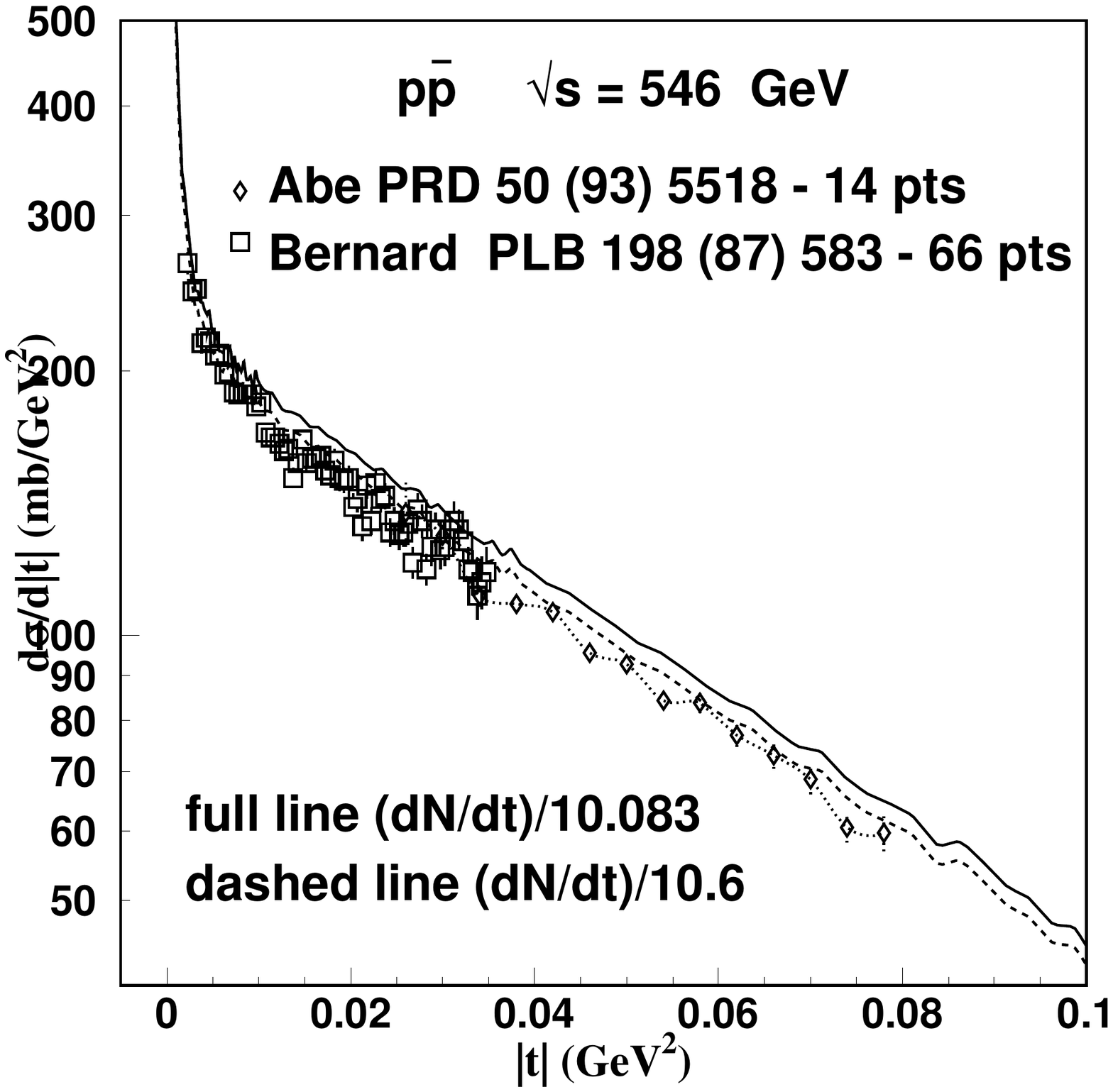}
\end{figure}

Observing locally the  comparison between the  prediction of 
$d\sigma/dt$  obtained from the event rate dN/dt by 
adjustment to Coulomb interference equations  and the data 
\cite{bozzo-I, bozzo-III}
we see that there is a discrepancy of a few percent. 
In Fig. \ref{normal_bozzo}  we show that the perfect 
matching is obtained with normalization factor 10.6,
 namely
  $$ \frac{d\sigma}{dt}  = \bigg(\frac{dN}{dt}\bigg)~ / ~ 10.6  ~ . $$
In Fig. \ref{bernard_abe} we show the differential cross-section
measured by Bernard et al. \cite{bernard}   and Abe et al. \cite{abe}
 which are compatible with each other, but do not match our 
 normalized solution for dN/dt. To match, we have to introduce an 
 arbitrary  normalization 
 factor 11.0 instead  of 10.083 that we have determined.   
 The  ratio $ \beta={B_R}/{B_I} $ in the range  
$$ 0.898 ~ \leq  ~ \beta=\frac{B_R}{B_I} ~  \leq  ~ 1.111  $$ 
leads to the same value 1.097 for $\chi^2$.  

To show the influence of value of  the ratio $ {B_R}/{B_I} $
beyond these limits , we have fitted with fixed 
    $$ \beta = {B_R}/{B_I}= 2 $$
obtaining  parameter values shown in  Table  \ref{parameters_541}.

 We have also calculated with fixed normalization factor 
$10.6$ , with results shown in the table. 
 \begin{center}
   \begin{table}
   \caption{ \label{parameters_541} Forward scattering parameters }
   \vspace{.5 cm}
   \begin{tabular}{|c|c|c|c|c|c|}
   \hline
 $\sigma ({\rm mb})$ & $ \rho $ & $B_I ({\rm GeV}^{-2})$ & $B_R ({\rm GeV}^{-2})$& normalization & $\chi^2$ \\
 \hline
$63.897 \pm 0.377$ & $0.172 \pm 0.009$ & $ 15.347 \pm 0.145 $ & $ 15.452 \pm 4.577$ & $10.083 \pm 0.135$ & $1.097 $ \\ 
$63.651 \pm 0.909$ & $0.1601 \pm 0.0168$ & $15.163 \pm 0.118$& $ 2 ~B_I ~({\rm fixed}) $ & $10.268 \pm 0.358$  & $ 1.115 $ \\
  \hline
 $62.688\pm 0.037$ & $0.1484 \pm 0.0025$ & $15.375 \pm 0.035$& $17.662\pm 1.441$ & $10.6 ~({\rm fixed})$ & $ 1.119 $  \\
 $62.842\pm 0.147$ & $0.1459 \pm 0.0070$ & $15.237 \pm 0.085$& $ 2~ B_I ~({\rm fixed})$ & $10.6 ~ ({\rm fixed})$ & $ 1.126 $ \\     
 \hline
   \end{tabular}
   \end{table}
   \end{center}

We see that the $\chi ^2$  values do not vary strongly, showing that the data can be described, within errors, by scattering parameters in  different ranges. 

\begin{figure}
 \caption {Correlations between the parameters $\rho$ and $\beta$ 
and between $\sigma$ and $\beta$ that lead to low values of $\chi^2$. 
In each case the other two parameters  are let free while
tables of $\chi ^2$ are built with specified values for the two 
plotted parameters. Also the normalization factor 10.083 is taken 
as fixed independently from PAW fitting program. }  
 \label{bolhas_541}
\includegraphics[height=10.0cm]{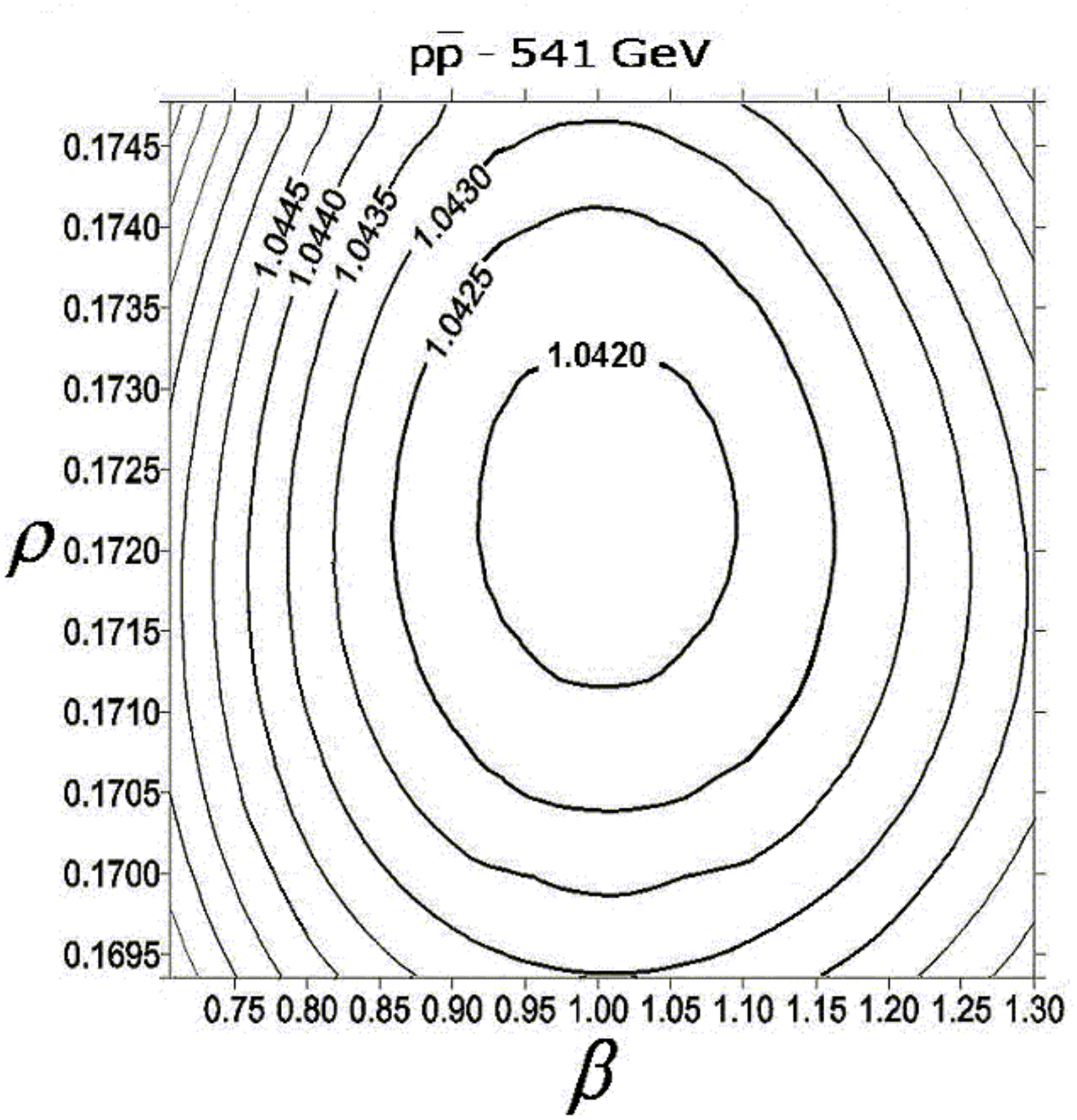}
\includegraphics[height=10.0cm]{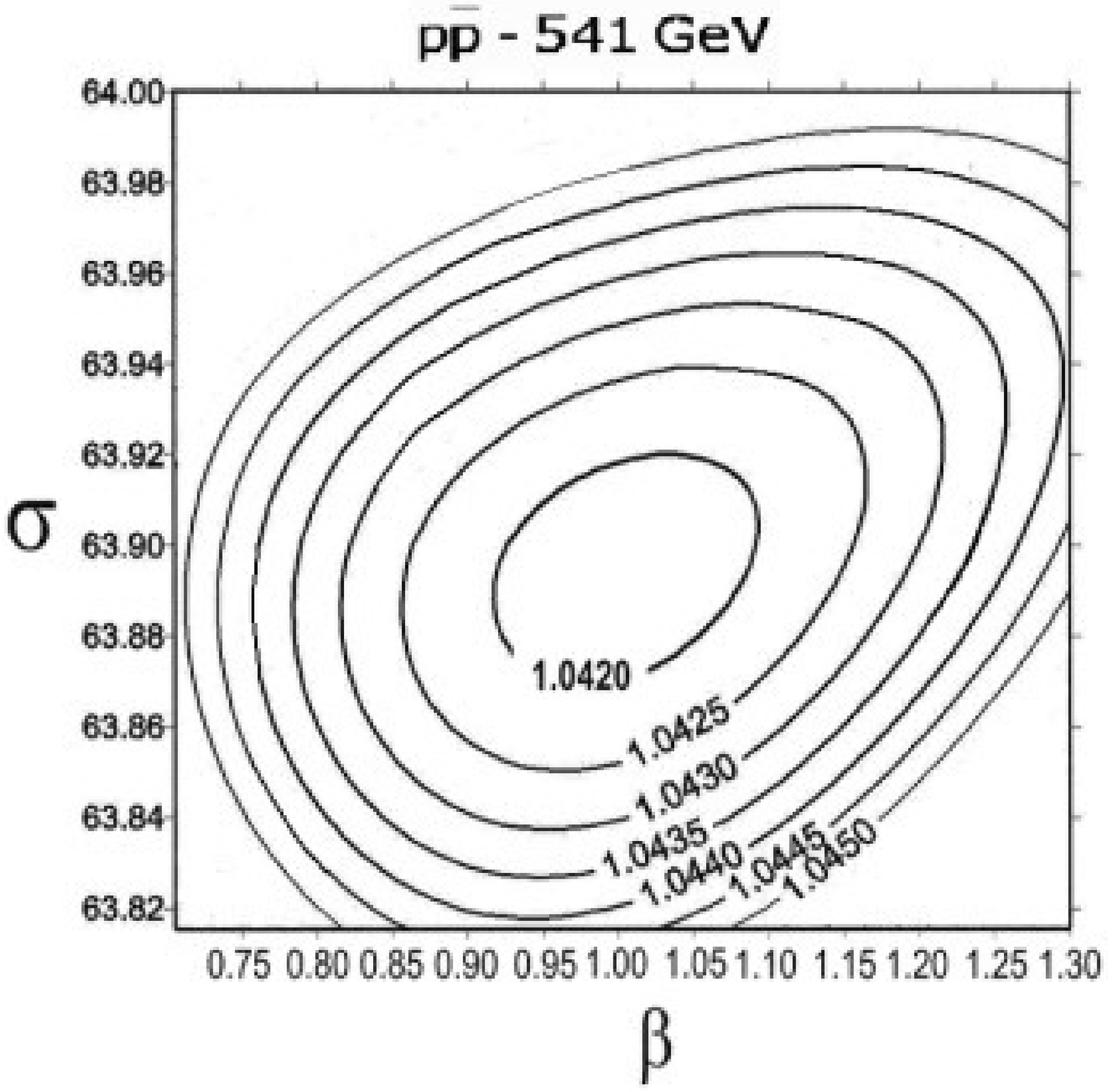}
\end{figure}

Correlations among parameters are shown in Fig. \ref{bolhas_541}  
with drawing of level lines determined for low $\chi^2 $ values.  

\begin{figure} 
 \caption {Data at 541 GeV selected and organized. The dotted line 
  is a fit described in our previous work \cite{pereira_ferreira}. } 
 \label{allpoints541}
\includegraphics[height=10cm]{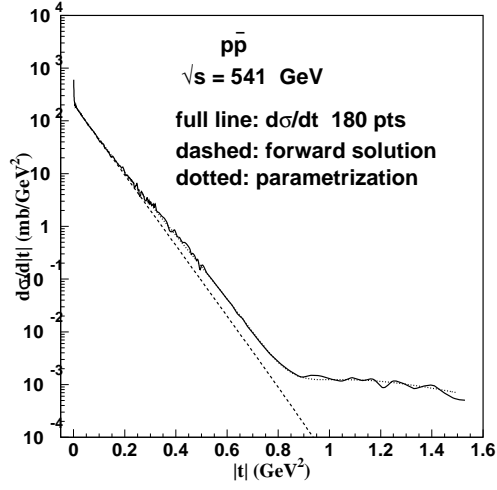}
\end{figure}

We have built a file with a continuous and non superposing set of 
points, being the first  59  points from  normalized dN/dt
(with normalization factor 10.6) and   121  points from  
Bozzo et al. The 180 points form a regular $|t|$  distribution, 
which we fit with formulae from our previous work \cite{pereira_ferreira}.
The results are shown in Fig. \ref{allpoints541}.
The parameter values obtained in this fitting are $\sigma = ~ 63.06 \pm 1.90 $ mb , 
$\rho = ~ 0.124 \pm 0.005 $, $B_I = ~ 13.88 \pm 0.42 ~ {\rm GeV}^{-2}$, 
$B_R = ~ 25.79 \pm 0.77 ~ {\rm GeV}^{-2}$, with $\chi^2=1.32$.   

%  \clearpage 

\subsection{ p$\bar {\rm p} $ scattering  at $\sqrt{s}$ = 1800 GeV}

\begin{figure}
 \caption {Data and results at 1800 GeV .  }  
 \label{data1800}
\includegraphics[height=9.0cm]{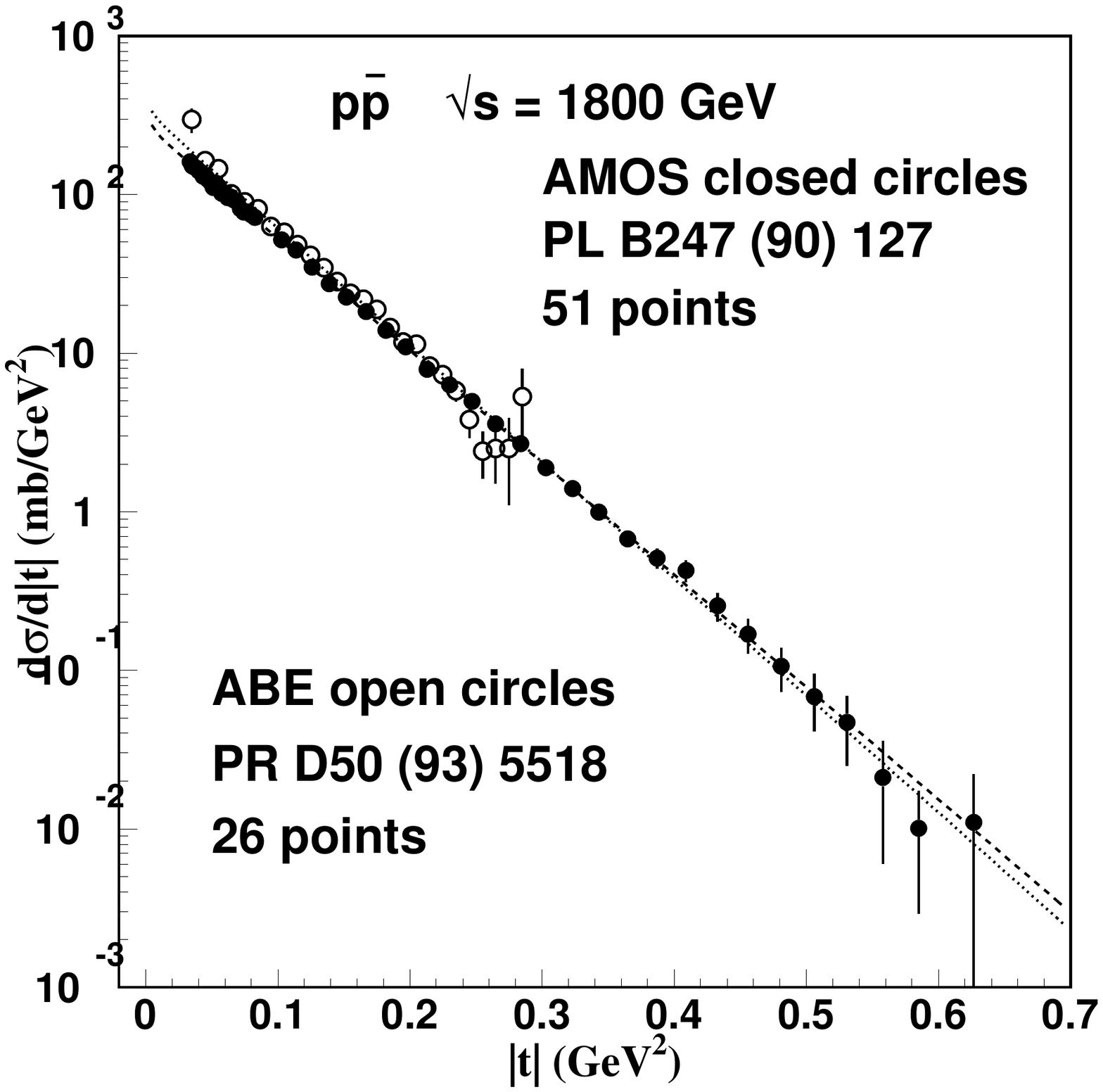}
\includegraphics[height=9.0cm]{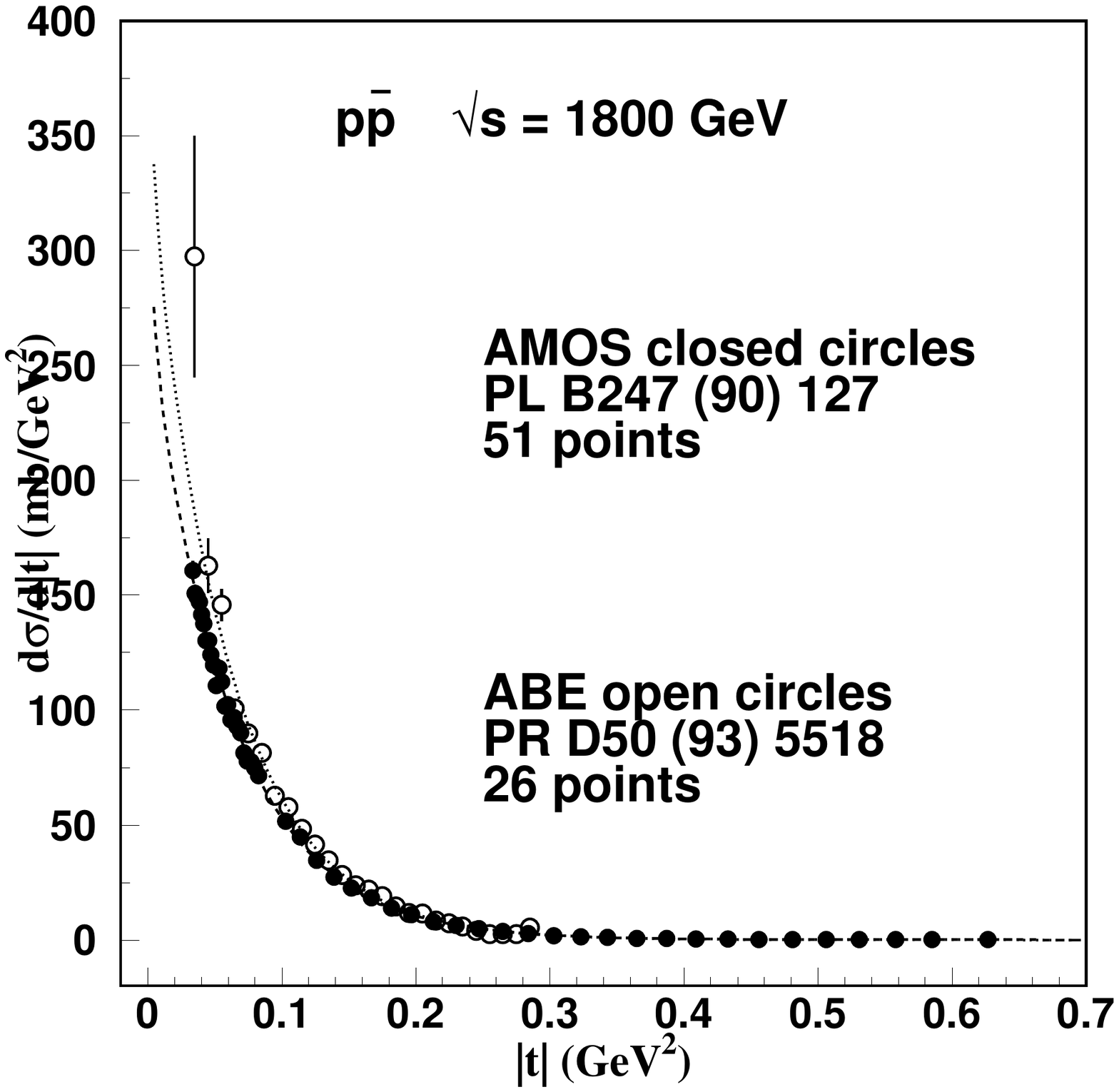}
\end{figure}

 Two independent experiments, both at Fermilab, measured
$d\sigma/dt$   for small $|t|$ at 1800 GeV, with a discrepancy 
of about 10 percent in the normalization for the total cross 
section. This is shown in Fig. \ref{data1800}.  
Our determination of parameters (with fixed $\rho=0.14$) is given in 
Table \ref{parameters_1800}.
Although this is not a close determination, due to large variation bars, 
it is interesting that the lowest $\chi^2$ are obtained with $B_R$ 
larger than $B_I$, for both experiments.. 
 \begin{center}
   \begin{table}
   \caption{ \label{parameters_1800} Forward scattering parameters at 1800 GeV }
   \vspace{.5 cm}
   \begin{tabular}{|c|c|c|c|c|c|}
   \hline
Experiment  &  $\sigma ({\rm mb})$ & $ \rho $ & $B_I ({\rm GeV}^{-2})$ & $B_R ({\rm GeV}^{-2})$&                                                              $\chi^2$ \\
 \hline
E710 &$72.748\pm 0.186$&$0.14$ (fixed)&$16.297 \pm 0.039 $&$115.57\pm 164.20 $&$0.6020 $ \\ 
E710 &$71.824 \pm 0.184$&$0.14$ (fixed)&$16.282 \pm 0.039 $&$  ~B_I ~({\rm fixed})$&$0.6060 $ \\
E710 &$72.651 \pm 0.186$&$1.0$ (fixed)&$16.284 \pm 0.039 $&$ 167.93\pm 48.561$&$0.5961 $ \\
  \hline
CDF  & $80.917\pm 0.436$& $0.14$(fixed)&$16.988\pm 0.087$& $72.006\pm 116.15 $ & $ 1.771 $  \\
CDF & $79.982\pm 0.432$& $0.14$(fixed)&$16.981\pm 0.087$& $  B_I ~({\rm fixed})$ & $ 1.775 $ \\  
CDF  & $80.159\pm 0.433$& $1.0$(fixed)&$16.865\pm 0.087$& $85.730\pm 16.937 $ & $ 1.705 $  \\
 \hline
   \end{tabular}
   \end{table}
   \end{center}
From the results shown in Table \ref{parameters_1800}, with large differences 
in $\chi^2$ values,  we learn that the data from the E710 
 experiment are more compatible with the forward scattering basic expression 
(\ref{dsigdt}) for $d\sigma/dt$ than the CDF data. Another important  observed 
result is that  increasing simultaneously the values $\rho$ and $B_R$ 
we may obtain the same, or lower, $\chi^2$ : the real amplitude at 
$|t|=0$ may be larger, but it decreases faster. It is as if for a given dataset
we should keep constant a product $\rho e^{-(B_R/2)|t|_{\rm eff}}$ , with 
a value $|t|_{\rm eff} $  that is effective for the Coulomb interference 
region. For the E710  data  we extract from the table 
that $|t|_{\rm eff} \approx 0.075 $ ; for the CDF basis the value is rather 
large $|t|_{\rm eff} \approx 0.28$ . We thus see that the values of $\rho$ 
and $B_R$ at 1800 GeV are not determined from the existing data 
in this analysis.

\begin{figure}
\caption {Data on p$\bar {\rm p}$ scattering at 1.8 and (preliminary) 
at 1.96 TeV. The lines in the LHS plot represent the fittings with the basic 
expression (\ref{dsigdt}) :  dashed for E710 and dotted line for CDF data. 
The  RHS plot shows the fitting (solid line) of all the data (dashed line)
with parameterization used in our previous work \cite {pereira_ferreira}  . }   
 \label{data1_8-1_96}
\includegraphics[height=9.0cm]{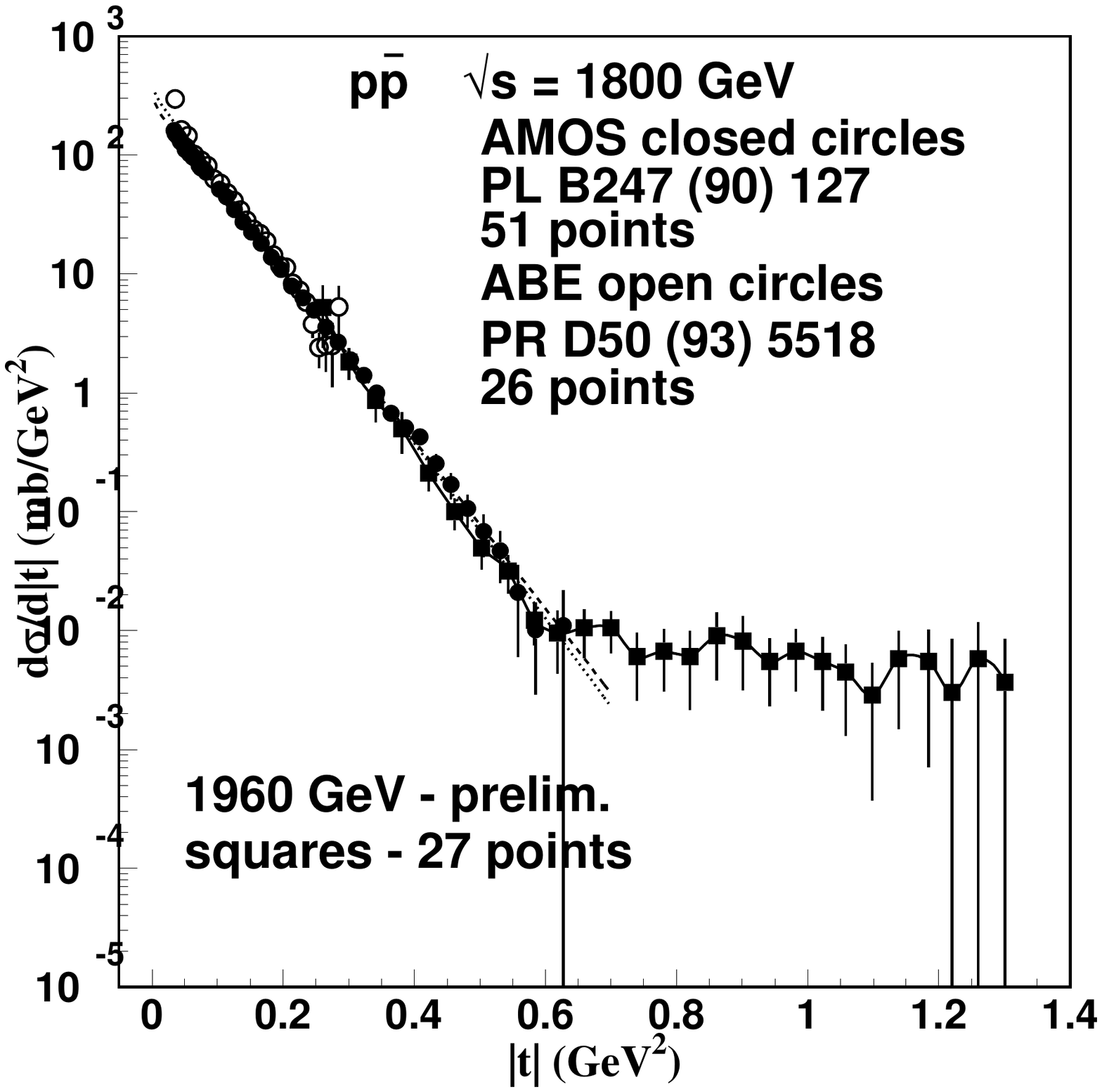}
\includegraphics[height=9.0cm]{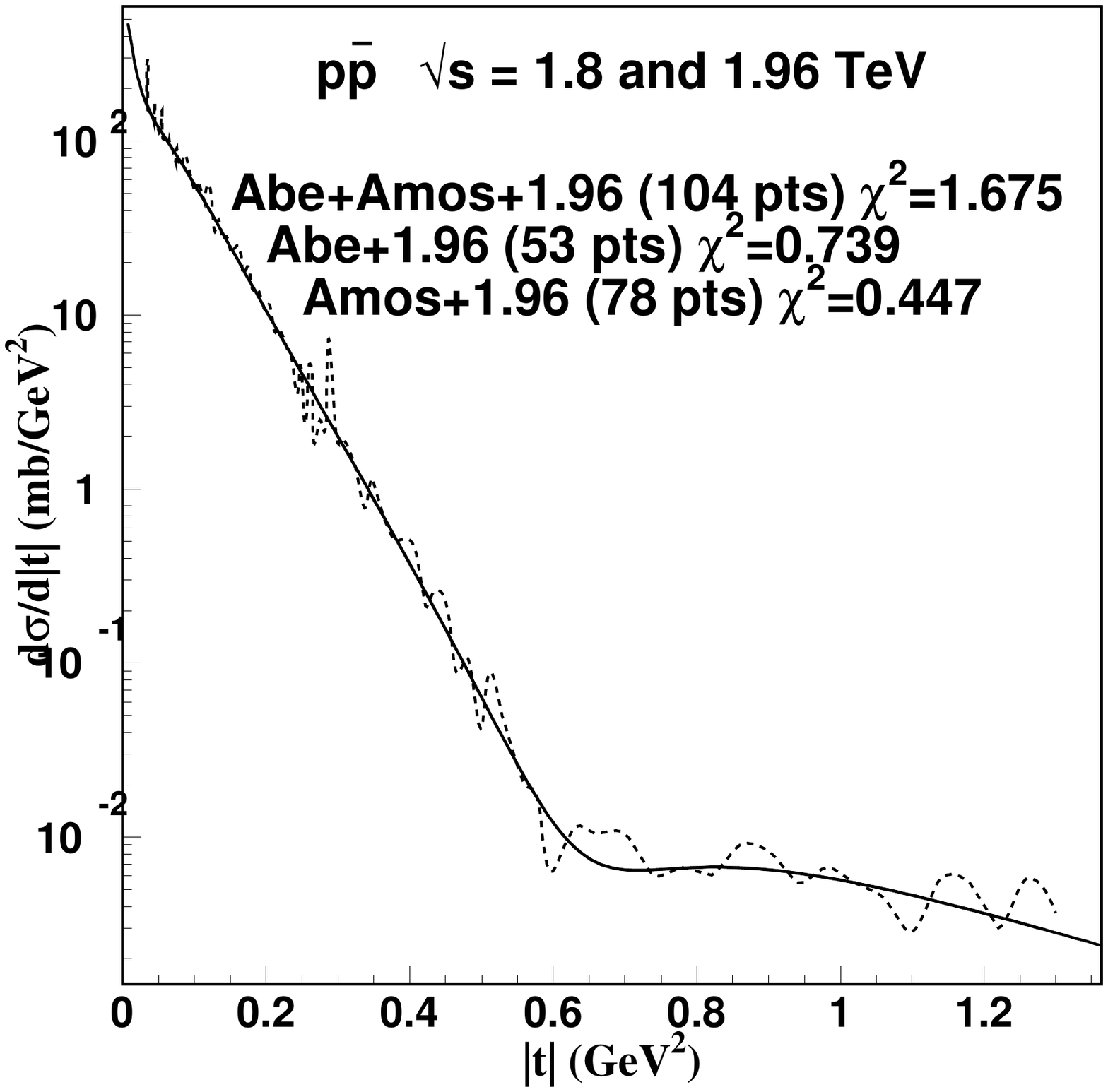}
\end{figure}

The Fermilab data on $d\sigma/dt$ at 1.96 TeV are now launched in preliminary 
 form, covering the $t|$ range from 0.26 to 1.30 $\GeV^2$. In Fig. 
\ref{data1_8-1_96} 
we show  these new data together with the old 1.8 TeV data. The LHS plot gives 
also the lines  representing the fittings of the E710 data (dashed line) and of 
the CDF data (dotted line) as informed in Table  \ref{parameters_1800},  
with fixed  $\rho=0.14$.  
 
The RHS figure shows the fitting (solid line) of all these data 
(marked by the interpolating dashed line) using our previous 
parameterization expressions \cite {pereira_ferreira} . We obtain 
$\chi^2=1.675$ for the whole lot of data (51+26+27=104 points).  
Taking only the E710 points together with the 1.96 TeV data 
(51+27=78 points) we obtain a beautiful representation with 
$\chi^2=0.45$; taking the CDF points together with the new 
 1.96 data (26+27=53 points) we obtain $\chi^2=0.74$. 
These results, that depend on our particular parameterization, but 
are meaningful  because they come from a search for continuity 
in the data, again seem to favor the $t$ dependence and normalization 
of the E710 data. 

\subsection{ Amplitudes } 

\begin{figure}
\caption {Amplitudes of p$\bar {\rm p}$ scattering at 541 and 1800 (including 
preliminary 1960) GeV normalized to 1 at $|t|=0$. The amplitudes are obtained  
with parameterization used in previous work \cite {pereira_ferreira}. 
Notice the linear scale in the LHS plot.
The zeros of the real and imaginary amplitudes at 1800/1960 GeV  occur 
respectively at about 0.05 and 0.68 $\GeV^2$ . }   
 \label{amplitudes}
\includegraphics[height=6.8cm]{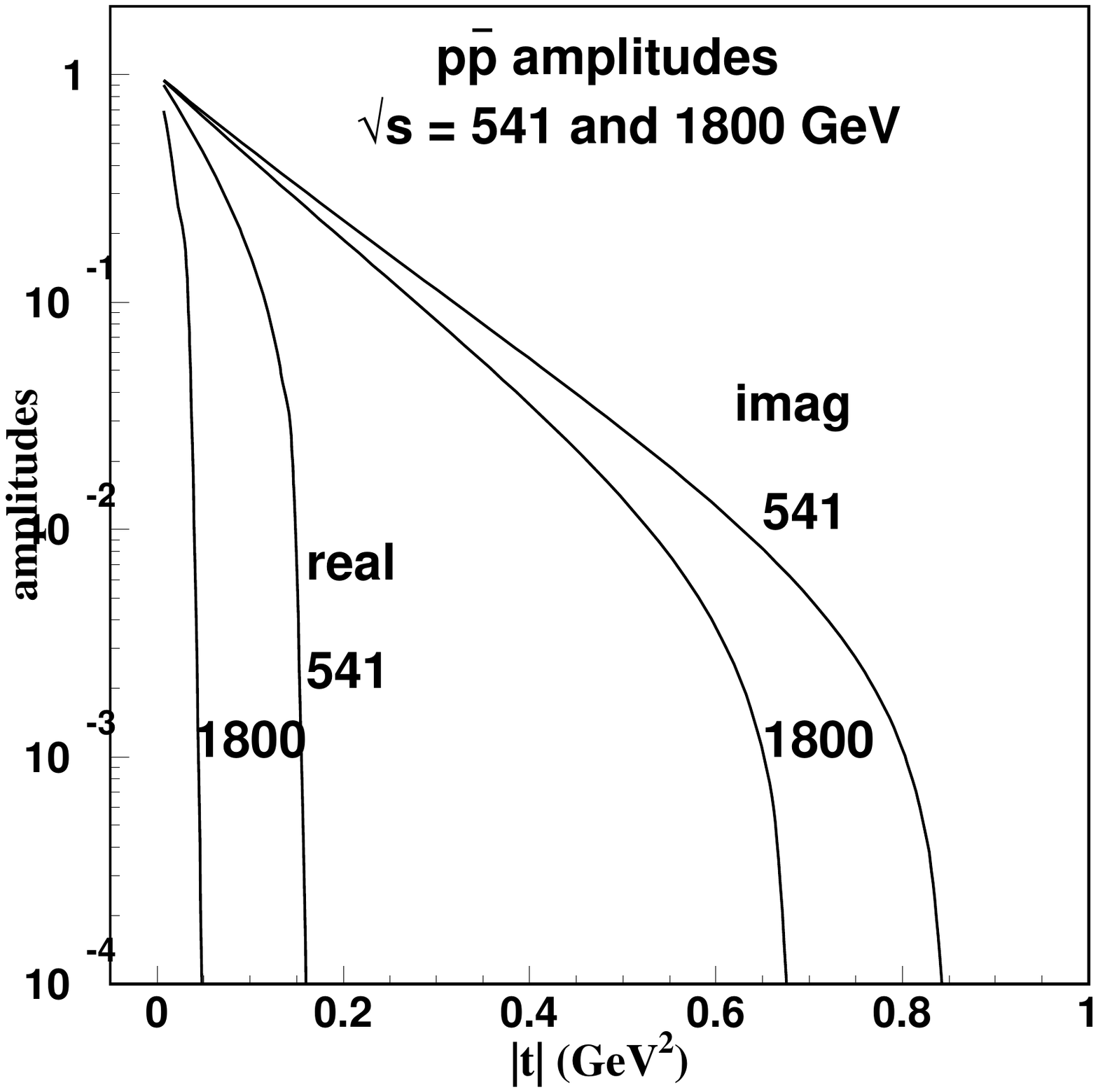}
\includegraphics[height=6.8cm]{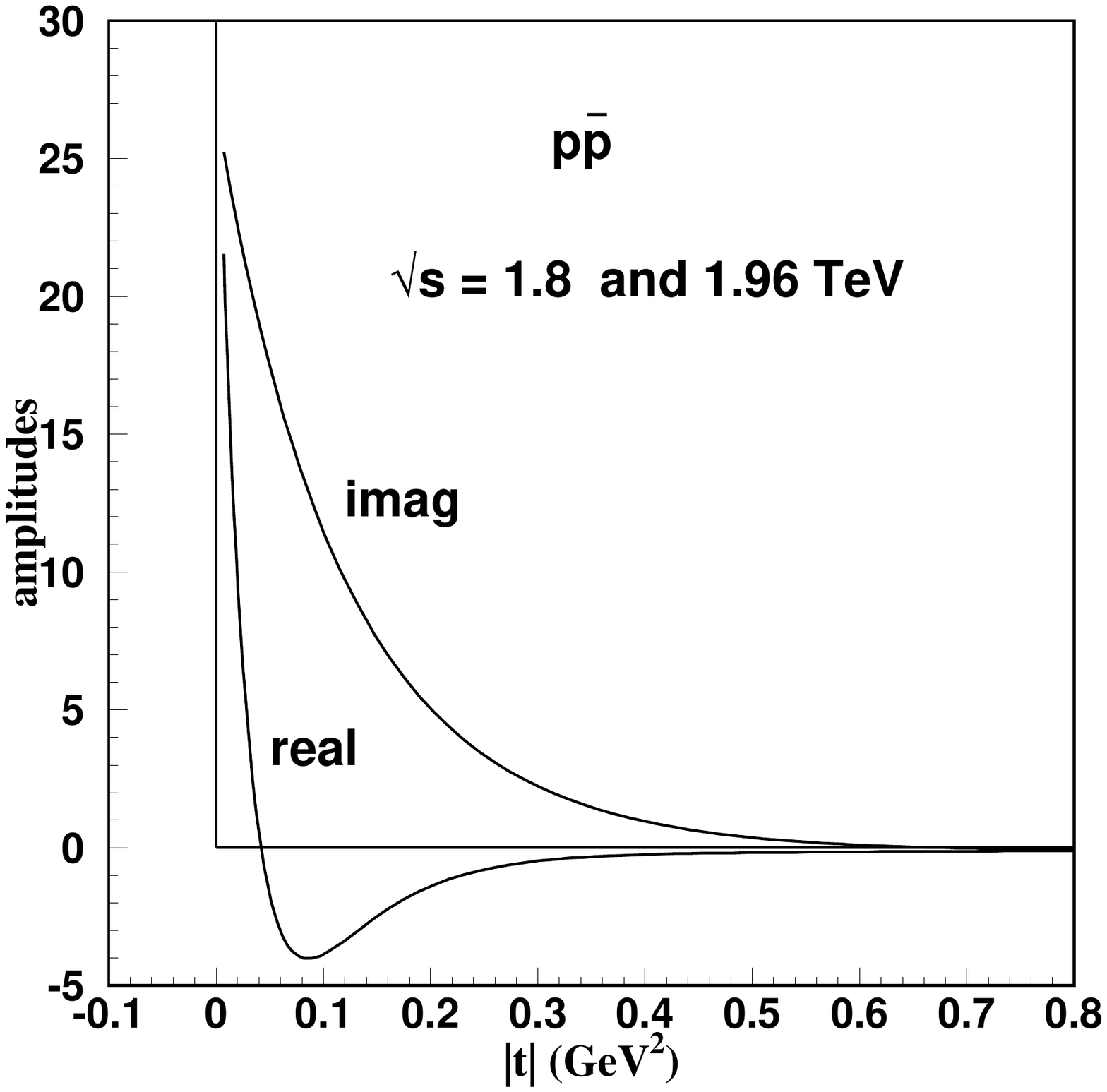}
\includegraphics[height=6.5cm]{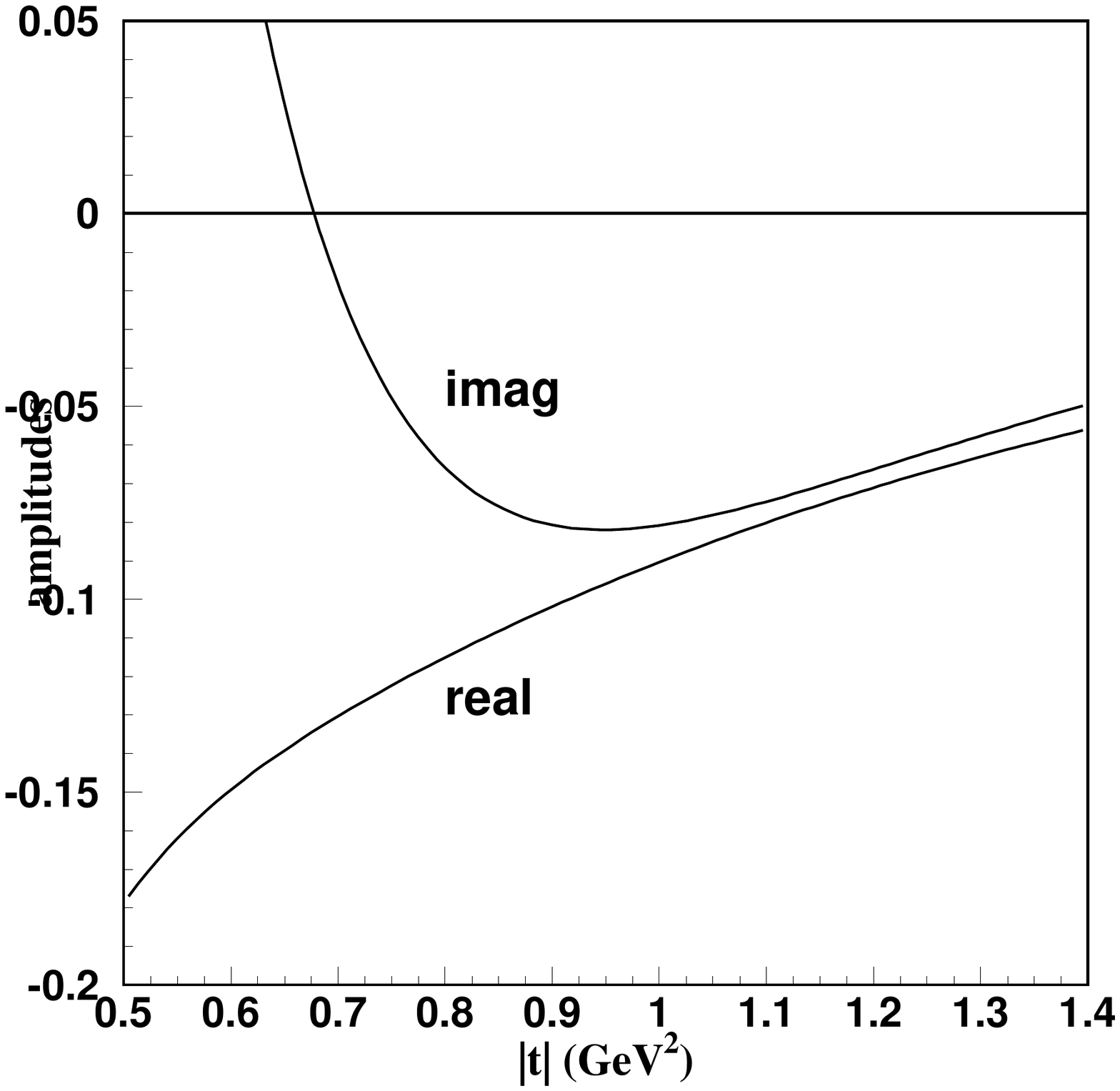}
 \end{figure}

The lines representing $d\sigma/dt$ in 
Figs. \ref{allpoints541} and \ref{data1_8-1_96} 
are obtained from real and imaginary amplitudes that are 
shown in Fig. \ref{amplitudes}, with their  
characteristic  slopes and zeros . 

Some typical values of parameters obtained in this analysis of the 
full $t$ dependence, that can be read off from the plots of the 
amplitudes,  are given in Table \ref{features}. These are possible, but 
not unique or   necessarily correct,  representations of the data.
 \begin{center}
   \begin{table}
   \caption{ \label{features} Quantities extracted from the $t$ dependence of the 
   amplitudes at 541/546 and 1800/1960 GeV. Remarks: (1)Abe+Amos+1.96~ ;~
 (2) Amos+1.96 ~; ~ (3) Abe+1.96 ~.}
   \vspace{.5 cm}
   \begin{tabular}{|c|c|c|c|c|c|c|c|}
   \hline
$\sqrt{s}$  & $\sigma $ & $\rho$ & $B_I $ & $B_R
      $ & $|t_0^R|$ & $|t_0^I|$ & $\chi^2$ \\
  (GeV)  &  ({\rm mb})  &        & $({\rm GeV}^{-2})$  & $({\rm GeV}^{-2})$ & $ ({\rm GeV}^{2})$       & $ ({\rm GeV}^{2})$   &       \\
\hline 
 541/546       & 63.05  & 0.12   & 13.88   & 25.79  & 0.16  & 0.85    & 1.32  \\
1800/1960~ (1) & 73.98  & 1.17   & 15.50   & 85.43  & 0.05  & 0.69    & 1.68   \\
1800/1960~ (2) & 73.95 & 0.75    & 15.47   & 84.12  & 0.05  & 0.69   & 0.45  \\
1800/1960~(3)  & 88.49 &  1.21  & 17.94  & 61.43   & 0.05  & 0.69   & 0.74  \\
  \hline
   \end{tabular}
   \end{table}
   \end{center}
The real amplitude shows a large slope and falls rapidly to zero at a value 
$|t_0^R|$ that is expected to approach the origin in the form \cite{pereira_ferreira}
\begin{equation}
\label{zero}
   |t_0^R|= \frac{1}{A+B \log{s}} ~, 
\end{equation}
following  Martin's discussion about the first zero of the real
amplitude \cite{Martin}.
Obviously, as $|t_0|$ approaches the origin, $B_R$ grows with the energy. 

The numbers indicate  much indetermination in this 
analysis of the old data. 
In particular, identification of the values of $\rho$ and $B_R$ 
seems to be difficult. 
Dispersion relations must be used as a guide towards 
the disentanglement, in particular with application of the 
derivative dispersion relations for slopes  \cite{EF2007}. 
We recall that the energy dependence in the measurements 
is crucial for the good use of dispersion relations, and we hope 
that LHC will  produce  diffractive  data at several energies ? 

 % ===============================================

 \section{remarks and conclusions} 

We have performed a detailed analysis of the data on differential elastic
cross-sections allowing for the freedom of different slopes for the real and
imaginary amplitudes, namely $B_{R}\neq B_{I}$, for the data points obtained
in the ISR/SPS(Cern) and Tevatron (Fermilab) experiments during the years
1960-1990. The principal conclusion of the present analysis is that different
values of the slopes, in particular the possibility of $B_{R}>B_{I}$ in 
accordance with the expectations  from Martin's theorem [2] and from 
dispersion relations [1], are perfectly consistent within the present errors
of experimental data.

Our investigation concerns the four quantities  relevant for the elastic
forward processes, namely, $\sigma,\rho,B_{R}$ and $B_{I}$. Studying the
behavior of $\chi^{2}$ values near its minimum and statistically equivalent
parameter ranges, we observe that the available data from Cern  and from
Fermilab for small $\left\vert t\right\vert $ at the energies $20-2000$ GeV
are not sufficient for a precise determination of these four parameters. With
real and imaginary amplitudes as independent quantities in the form 
 $\alpha ~ e^{\beta t}$,  the $\chi^{2}$ analysis clearly shows a strong 
correlation among the parameters, exhibiting  a very large  valley in 
$\chi^{2}$ surface in the parameter space. This in particular leaves  
large ambiguities in the determination for the weaker real part.

It is interesting to note that detailed behavior in the $\chi^2$  surface
in the parameter space depends  very much on the chosen data sets. For
example, the minimum of  $\chi^2$/degrees-of-freedom varies from 0.1 to 1.3
in different energies. On the other hand, the variation of $\chi^2$ 
for appreciable changes of some parameters is  less than 0.1 
percent, in general indicating that their precise determination
is not possible within the existing experimental situation. 

On the other hand, it is worthwhile to mention that, if we have sufficient 
number of data points in this region, we do not need the absolute 
normalization of the luminosity, since the Coulomb interference can determine 
correctly the absolute value of cross section as a free parameter, as was 
shown in this work for the 541 GeV case. Of course, this method requires 
that the low $\left\vert t \right\vert $ data be very accurate. It is 
interesting to note that tests of
this of Coulomb interference method, introducing a new free normalization
parameter into the data set, are found to be compatible in both E-710 and
E-741 Fermilab  data  at $1800$ GeV, while the difference in the evaluation 
of the total cross section remains.

As mentioned before, the quantitative analysis for the disentaglement of 
the real and imaginary parts of
complex amplitude at small $\left\vert t\right\vert $ cannot be made with
confidence with the data available up to now. Tests of quantities like the
position of the zero of the real amplitude are not safe in these
conditions. Thus model independent values for the four parameters cannot be
obtained accurately. 

Usually, the models suited for pp dynamics aimed to cover an overall large
$\left\vert t\right\vert $ region, with correct description of dip and tail in
$d\sigma/dt$,  are not sensitive enough  to the details of the behavior of
scattering amplitudes for very small $\left\vert t\right\vert$ .  In this
region, the behavior of the amplitude maybe very sensitive to specific
dynamical influences of the non-pertubative QCD dynamics. In particular, the
forward scattering amplitude is directly related with the proton structure,
and intimately related with the parton distribution function at small $x$ and
saturation problems. 

As a complimentary study, we also performed the analysis of scattering
amplitude for a larger $\left\vert t\right\vert $ domain as proposed
in \cite{pereira_ferreira}. We observe that such analysis, consistent with 
the general
structure of the scattering amplitudes, such as positions of zeros and dips,
clearly shows the necessity of the very distinct values of the slope
parameters of real and imaginary amplitudes, corroborating with the results
obtained here from the analysis in small $\left\vert t\right\vert $ 
regions. In terms of scattering amplitude, this different behaviors of real
and imaginary part is crucial and will be fundamental to understand the
mechanism of elastic scattering amplitude.

Of course, the separation of complex amplitudes will never be a very easy task 
only from the measured scattering experiments, and it will be unavoidable 
to make use of other theoretical tools, such as dispersion relations. 
Particular attention
must be given to the development and application of dispersion relations for
slope parameters [1]. The knowledge of the energy dependence is crucial for 
the application of these tools in practice, and it is extremely interesting 
if an energy scan program is also included in the first phase of the LHC 
operation of p+p collisions.

Although no new or better method has been introduced in the present work,
our precise analysis has revealed the existence of correlations and
related uncertainties of the behavior of the scattering amplitude at small
$\left\vert t\right\vert$. For quantitative determinations, more precise 
and numerous data points are necessary.

We expect to have a different situation in the future experiments from RHIC
and LHC, with much better statistics and accuracy in the measurements of
scattering data in the Coulomb interference region, together with a systematic 
energy  scan program. 

\begin{acknowledgements}
 The authors wish to thank  
CNPq (Brazil), FAPERJ (Brazil)and PRONEX (Brazil) for general support
of their research work, including  research fellowships and grants. 
  \end{acknowledgements}
 
% ===============================================
\section {Appendix: The calculation of the Coulomb phase} 

Here we give details of the evaluation of the phase of West and Yennie 
given by eq. (\ref{WYphase}) in the case where we let $B_R \neq B_I$ .

  After eq. (\ref{relation1}) we define 
\begin{eqnarray}
\label {def_G}
G_R=\frac{c}{c+i}=\frac{c(c-i)}{c^2+1} \nonumber \\
G_I=\frac{1}{c+i}=\frac{(c-i)}{c^2+1}  ~ ,
\end{eqnarray}
and the integral in eq.(\ref{WYphase})  is written 
\begin{eqnarray}
\label{WY_integral}
&& \int_{-4p^2}^{0}
\frac{dt^\prime}{|t^\prime-t|}\bigg[1-\frac{F^N(s,t^\prime)}{F^N(s,t)}\bigg]
\nonumber \\
% =\int_{-4p^2}^{0}
% \frac{dt^\prime}{|t^\prime-t|}\bigg[G_R+iG_I-G_R e^{B_R(t^\prime-t)/2}-i G_I
% e^{B_I(t^\prime-t)/2}  \bigg] \nonumber \\
&=&  G_R  \int_{-4p^2}^{0}
\frac{dt^\prime}{|t^\prime-t|}\bigg[1-e^{B_R(t^\prime-t)/2}\bigg]
+ iG_I  \int_{-4p^2}^{0}
\frac{dt^\prime}{|t^\prime-t|}\bigg[1-e^{B_I(t^\prime-t)/2}\bigg] ~ , 
\end{eqnarray}
and then 
\begin{eqnarray}
\label{WYphase_2} 
\Phi(s,t)&=&(-/+)\Bigg[\ln(-\frac{t}{s})+
 G_R  \int_{-4p^2}^{0}
\frac{dt^\prime}{|t^\prime-t|}\bigg[1-e^{B_R(t^\prime-t)/2}\bigg] \nonumber \\
&+& iG_I  \int_{-4p^2}^{0}
\frac{dt^\prime}{|t^\prime-t|}\bigg[1-e^{B_I(t^\prime-t)/2}\bigg]
\Bigg] ~ . 
\end{eqnarray}
We note that both integrals are of the form
\begin{equation}
\label{int_form}
I(B)=\int_{-4p^2}^{0}
\frac{dt^\prime}{|t^\prime-t|}\bigg[1-e^{B(t^\prime-t)/2}\bigg] ~  
\end{equation}
which has been studied by  V. Kundr\'at  and M. Lokajicek \cite{KL}.
With $x=t^\prime-t$   and $y=Bx/2$, we have 
\begin{eqnarray}
I(B) &=& \int_{-4p^2}^{0}
\frac{dx}{|x|}\bigg[1-e^{B x/2}\bigg]
=\int_{-B(4p^2+t)/2}^{-Bt/2}
\frac{dy}{|y|}\bigg[1-e^y\bigg]  \nonumber \\
&=&
\int_{-B(4p^2+t)/2}^{0}
\frac{dy}{|y|}\bigg[1-e^y\bigg]
+ \int_{0}^{-Bt/2} \frac{dy}{|y|}\bigg[1-e^y\bigg] \nonumber \\
&=& \int_{0}^{B(4p^2+t)/2} 
\frac{dy}{|y|}\bigg[1-e^{-y}\bigg]
- \int_{0}^{-Bt/2} \frac{dy}{|y|}\bigg[e^y-1\bigg]  
\end{eqnarray}
These expressions can be written in terms of exponential integrals, 
as can be seen in the  Handbook of Mathematical Functions of 
M. Abramowitz and L.A. Stegun \cite{abramowitz} as 
\begin{equation}
\label{first}
\int_{0}^{B(4p^2+t)/2}  
\frac{dy}{|y|}\bigg[1-e^{-y}\bigg]= E_1\bigg[\frac{B}{2}\bigg(4 p^2+t\bigg)\bigg]
+\ln \bigg[\frac{B}{2}\bigg(4 p^2+t\bigg)\bigg]+\gamma
\end{equation}   
and 
\begin{equation}
\label{second}
\int_{0}^{-Bt/2} \frac{dy}{|y|}\bigg[e^y-1\bigg]  
=E_i\bigg(-\frac{B}{2}t\bigg)-\ln\bigg(-\frac{B}{2}t\bigg) - \gamma
\end{equation}\
where $\gamma = 0.5772 $ is the Euler constant.
 
Thus the integrals that appear in  eq. (\ref{WY_integral})  have the functional form 
\begin{equation}
\label{functional_form}
I(B)=   E_1\big[\frac{B}{2}\bigg(4p^2+t\bigg)\big] 
-E_i \big[-\frac{Bt}{2} \big] +\ln  \big[\frac{B}{2}\bigg(4p^2+t\bigg)\big]
-\ln  \big[-\frac{Bt}{2} \big] + 2 \gamma  ~ , 
\end{equation}
and the phase can be written
\begin{equation}
\Phi(s,t)=(-/+)\Bigg[\ln \bigg(-\frac{t}{s} \bigg) 
     +G_R ~ I(B_R) + iG_I ~ I(B_I) \Bigg]      
\end{equation}
or 
\begin{equation}
\Phi(s,t)=(-/+)\Bigg[\ln \bigg(-\frac{t}{s} \bigg) 
+ \frac{1}{c^2+1}\bigg[ c^2 I(B_R)+I(B_I) \bigg]
+ i\frac{c}{c^2+1}\bigg[I(B_I)-I(B_R) \bigg]
\Bigg] ~ .      
\end{equation}
The real part of the phase is taken into eq. ( \ref{dsigdt1} ) . 

% ==============================================

\end{document}